\documentclass[twocolumn,trackchanges]{aastex701}

\usepackage{amsmath}
\DeclareMathOperator*{\med}{med}
\DeclareMathOperator*{\std}{std}

\graphicspath{{./}{figures/}}

\begin{document}

\title{Interference Meets Inference: Bayesian Time-Series Modeling of Radio-Frequency Interference}

\author[orcid=0009-0004-0966-5874, gname=Jade, sname=Ducharme]{Jade M. Ducharme}
\affiliation{Department of Physics, Brown University, Providence, RI, USA}
\email[show]{jade\_ducharme@brown.edu}  

\author[orcid=0009-0007-8793-6969,gname=Andrei, sname=Li]{Andrei Li} 
\affiliation{Department of Physics and McGill Space Institute, McGill University, 3600 University Street, Montreal, QC H3A 2T8, Canada}
\email{andrei.li@mail.mcgill.ca}

\author[orcid=0000-0001-7716-9312, gname=Michael ,sname=Wilensky]{Michael J. Wilensky}
\altaffiliation{CITA National Fellow}
\altaffiliation{TSI Post-Doctoral Fellow}
\affiliation{Department of Physics and McGill Space Institute, McGill University, 3600 University Street, Montreal, QC H3A 2T8, Canada}
\email{michael.wilensky@mcgill.ca}

\author[orcid=0000-0001-6876-0928, sname=Liu, gname=Adrian]{Adrian Liu}
\affiliation{Department of Physics and McGill Space Institute, McGill University, 3600 University Street, Montreal, QC H3A 2T8, Canada}
\email{adrian.liu2@mcgill.ca}

\begin{abstract}

Radio-frequency interference (RFI) remains a major challenge for modern radio astronomy experiments. In this work, we cast RFI detection as a one-dimensional time-series anomaly-detection problem and develop a probabilistic mixture-model framework for separating a smoothly varying astronomical background from anomalous contamination. The model jointly describes the clean and contaminated components and assigns each time sample a posterior probability of belonging to the RFI state, rather than relying solely on binary flags. This probabilistic formulation provides a measure of classification confidence, enables uncertainty propagation into derived downstream statistics, and offers additional information for investigating ambiguous events. We apply the framework to observations from the Murchison Widefield Array collected in 2014, producing ``soft" classification labels and seasonal RFI trends. We perform a parallel analysis using the Sky-Subtracted Incoherent Noise Spectra software pipeline (\textsc{SSINS}), which produces ``hard" classification labels. Overall, the mixture model provides similar classification labels but offers a more comprehensive framework by providing a complementary probabilistic description of RFI contamination and its uncertainty.

\end{abstract}



\section{Introduction}

Radio-frequency interference (RFI) is an increasingly pressing problem for radio astronomy operations. High-precision applications of radio astronomy, such as those targeting cosmological measurements, are particularly impacted by the rising number of anthropogenic RFI contaminants  \citep{Wilensky_2020, Wilensky_2023}. Common sources include direct satellite emissions \citep{DiVruno_2023, Grigg_2023, starlink3, Grigg_2025}, terrestrial signals reflected by aircraft \citep{Ducharme_Pober_2025} or satellites \citep{prabu2023}, and ground-based transmissions propagated over long distances via tropospheric ducting \citep{indermuehle2026}.

To prevent such ubiquitous contamination from propagating through to a downstream scientific analysis, preliminary \textit{flagging} treatments are typically applied to astronomical data. These aim to identify contaminated samples, which are then excluded from subsequent processing and analysis.

Existing flagging algorithms typically produce categorical decisions, classifying each sample as either clean or contaminated \citep{AOFlagger, Wilensky_2019, Kunicki_Pober_2024}. Such decisions necessarily impose a threshold between the two classes, which can lead to both under-flagging (in which residual RFI remains in the data) and over-flagging (in which otherwise usable data are unnecessarily discarded). A flagging framework of particular relevance to this study is the Sky-Subtracted Incoherent Noise Spectra (\textsc{SSINS}; \citealt{Wilensky_2019}), developed for the detection of faint RFI in interferometric data. \textsc{SSINS} has been used extensively for RFI identification in observations from the Murchison Widefield Array (MWA; \citealt{mwa1, mwa2}), including analyses targeting reionization \citep{Barry2019, Li_2019, Nunhokee_2025}.

While such categorical flagging methods provide an effective means of identifying and excising contaminated data, the final binary classification discards information about the uncertainty associated with each decision. Retaining this information is potentially valuable both for distinguishing ambiguous samples near the boundary between clean and contaminated data and for characterizing the underlying RFI population itself. This motivates a probabilistic treatment of the flagging problem. Related Bayesian approaches have been explored by \citet{Leeney_2023}, who modeled RFI contamination directly within the likelihood, and more recently by \citet{Leeney_2026}, who applied a similar approach to automated data curation for the REACH\footnote{\url{https://www.reachtelescope.org/}} experiment.

In this work, we develop a Bayesian mixture-model framework for characterizing and flagging RFI in time-series data. Our approach jointly models the smoothly varying background and the distributions of clean and RFI-contaminated residuals. Rather than assigning a hard categorical flag to each time sample, our method estimates its posterior probability of belonging to either class. The framework also infers population-level parameters describing the clean and contamination components, enabling several forms of statistical analysis of the RFI, as we shall demonstrate. We apply the framework to native \textsc{SSINS} data products from MWA observations, allowing a direct comparison with an established flagging approach.

Although demonstrated here using observations from the MWA, the proposed methodology is broadly applicable to other telescopes, including non-interferometric instruments, as well as to anomaly detection in time-series data more generally. We therefore begin with a general formulation of the mixture modeling framework in Section~\ref{sec:modeling_framework}. In Section~\ref{sec:preprocessing}, we detail the pre-processing steps applied to our interferometric input data to produce the one-dimensional representation expected by the mixture model. In Section~\ref{sec:implementation}, we describe the practical implementation of the inference procedure. In Section~\ref{sec:real_analysis}, we apply the framework to the 2014 observing season from the MWA's Phase I Epoch of Reionization (EoR) highband data. We compare the resulting classifications with flags obtained via \textsc{SSINS}, while also deriving season-level conclusions about the RFI environment. We discuss limitations and potential future improvements in Section~\ref{sec:limitations} and present our key findings and conclusions in Section~\ref{sec:conclusion}. 

\section{Modeling framework} \label{sec:modeling_framework}

Our modeling framework expects input data in the form of a one-dimensional time series. Although the interferometric visibilities used in our worked example are recorded as a function of time, they also span several additional dimensions such as frequency, baseline, and polarization. The baseline axis is typically the largest of these dimensions, with each baseline corresponding to the correlation between a pair of antennas in the interferometer. Section~\ref{sec:preprocessing} describes how we process and compress the raw input visibilities into the time-series representation required by the implementation presented here. The restriction to one-dimensional time series is a simplifying choice of the present implementation rather than a fundamental limitation of the mixture-model approach, which could be extended to retain frequency- or antenna-dependent information in future work.

\begin{figure}
    \centering
    \includegraphics[width=1.0\linewidth]{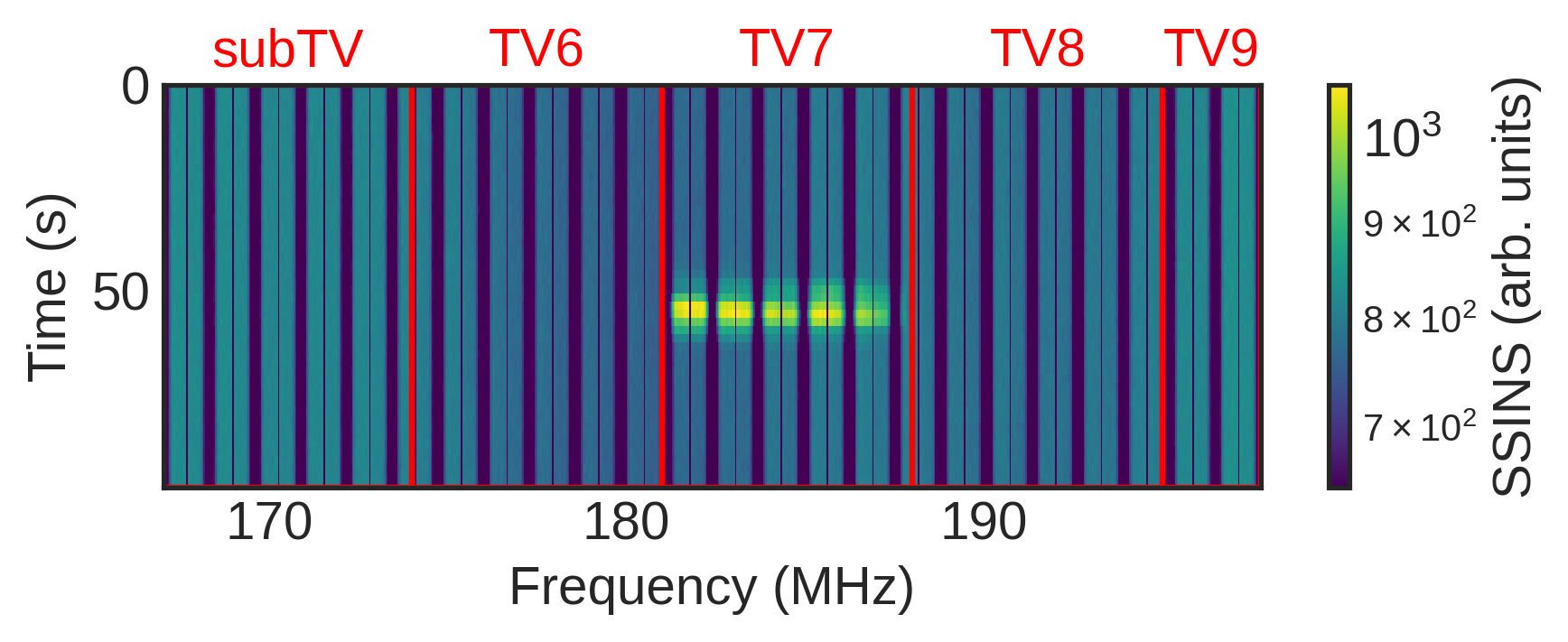}
    \caption{A waterfall plot for the two-minute MWA observation with ID 1089660568, corresponding to a short segment of an observing night, after processing with \textsc{SSINS} (see Section~\ref{sec:ssins}). A clear instance of RFI contamination in the TV7 band is seen around 50 seconds into the observation. The regular vertical striping arises from aliasing at the boundaries of the MWA's coarse frequency channels \citep{Li_2019}. These corrupted frequencies are flagged by \textsc{SSINS} and not used in the rest of our analysis. The five frequency sub-bands of interest for this study are outlined in red (see Section~\ref{sec:data_compression}).}
    \label{fig:waterfall}
\end{figure}

The first pre-processing stage involves applying the \textsc{SSINS} framework, as described in detail in Section~\ref{sec:ssins}. During this first stage, the data are compressed along the baseline axis, yielding the SSINS data product that remains a function of time, frequency, and polarization. For a fixed polarization, the SSINS can be visualized as a two-dimensional time-frequency waterfall plot, an example of which is shown in Figure~\ref{fig:waterfall}. In this space, RFI appears as a positive excess in the SSINS, as illustrated by the bright feature spanning 181--188 MHz at approximately 50 seconds. Further compression steps detailed in Section~\ref{sec:data_compression} reduce the input data to the required one-dimensional representation.

Representing the data as time series rather than waterfalls substantially reduces storage and computational requirements. More broadly, because the modeling framework operates on generic time-series inputs, it can be applied beyond interferometric data to a wide range of anomaly detection problems.

We therefore introduce our modeling framework as a tool generally suitable for \textit{time-series anomaly detection}. We model the input data as comprising three main components: a smoothly varying clean background component, noise, and an anomalous component corresponding to RFI. Jointly modeling all three components simultaneously helps prevent RFI from biasing the model of the clean background.

In the following sub-sections, we detail how each component is modeled (Sections~\ref{sec:background}--\ref{sec:rfi}), presenting the assumptions and choices made for our specific use case before introducing a generalizable joint mixture model in Section~\ref{sec:mixture_model}. In Section~\ref{sec:classification}, we introduce a classification scheme based on the joint model that quantifies the confidence that each input sample is either clean or contaminated by RFI.

\subsection{Smooth background}\label{sec:background}

In our model, we assume a smoothly varying background. Variations in the background are taken to occur on longer timescales and have smaller amplitudes than anomalous events. This difference in timescale is what allows us to differentiate between the smooth uncontaminated background and anomalous components. For our case study, these components correspond to uncalibrated interferometric data in the absence of RFI and to RFI contamination which is present on top of the background.

The smoothly varying background results from temperature-dependent antenna gains in the uncalibrated data. Among other factors, antenna gain is inversely correlated with physical antenna temperature. Smooth temperature fluctuations arising from the antennas' refrigeration cycle thus lead to smooth fluctuations in receiver gains \citep{barry_2018}. This translates in turn to smooth fluctuations in the resulting time series. The details of this are discussed more in depth in Section~\ref{sec:ssins}.

The smooth background consists of percentage-level fluctuations about a constant offset. We choose to model it using a Legendre polynomial basis. While other bases could provide similarly suitable representations of the smooth background, we adopted Legendre polynomials for their simplicity and ability to capture the expected smooth fluctuations. The background is modeled on a per-night basis:
\begin{equation}\label{eq:legendre_bg}
    \mathbf{b}^\mathrm{n} = \beta_0^\mathrm{n} \mathbf{L_0} + \beta_1^\mathrm{n} \mathbf{L_1} + \beta_2^\mathrm{n} \mathbf{L_2} + ...
\end{equation}
where $\mathbf{L_\ell}$ denotes the $\ell$-th Legendre polynomial evaluated at the $T_\mathrm{n}$ time samples for a given night, $\beta_\ell^\mathrm{n}$ are the corresponding scalar Legendre coefficients, and n indexes the nights. The number of Legendre polynomials employed for the fit was truncated at $L=24$ to prevent overfitting. Motivation behind this choice of term number is discussed in Section~\ref{sec:model-tailoring}.

The first Legendre polynomial is the constant $\mathbf{L_0} = \mathbf{1}$. In our model, the first Legendre mode $\ell=0$ thus corresponds to the constant offset for that night. Higher order Legendre modes in turn correspond to fluctuations from this constant offset. We model each $\beta_\ell^\mathrm{n}$ as being drawn from a normal distribution:
\begin{equation}
    \beta_\ell^\mathrm{n} \sim \mathcal{N}(\mu_{\beta_\ell}, \sigma_{\beta_\ell}),
\end{equation}
where $\mu_{\beta_\ell}$ and $\sigma_{\beta_\ell}$ are, respectively, the population mean and standard deviation of the $\ell$-th Legendre coefficient across nights. For $\ell>0$, these are given weakly informative priors (see Section~\ref{sec:specific_priors}). Meanwhile, the $\ell=0$ mode depends strongly on the specific night considered. As such, its coefficient priors are computed per-night according to
\begin{equation}
    \begin{split}
        \mu_{\beta_0}^\mathrm{n} &= \med\,\mathbf{y}^\mathrm{n},\\
        \sigma_{\beta_0}^\mathrm{n} &= 5\,\std\,\mathbf{y}^\mathrm{n},
    \end{split}
\end{equation}
where $\mathbf{y}^\mathrm{n}$ is the input time series for the n-th night. The median and standard deviation are taken across the time axis. That is, we center each nightly $\ell=0$ coefficient prior mean on the median data value for that night, and give it flexibility by setting $\sigma_{\beta_0}^\mathrm{n}$ to five times the standard deviation of data on that night.

\subsection{Estimating the noise}
\label{sec:noise}

The noise properties of the data for our use case can be estimated from theoretical principles based on how the data product is formed. As noted at the start of the section, the input visibilities are first processed via the \textsc{SSINS} software, which we describe in detail in Section~\ref{sec:ssins}. To estimate the noise standard deviation, we use a modified form of the equation from \cite{Wilensky_2019}
\begin{equation}\label{eq:noise}
    \sigma_\text{noise}^\mathrm{n} = \sqrt{\frac{C \hat{\mu}^2}{N_\text{bl}N_\text{chan}}},
\end{equation}
where
\begin{equation}
    C = \frac{4}{\pi} - 1
\end{equation}
is the ratio of the Rayleigh variance to the square of its mean (estimated here by $\hat{\mu}$). The Rayleigh distribution is relevant since in the absence of RFI, our data are calculated from a deep average of amplitudes of (independent and approximately identically distributed) circular complex Gaussian noise. These amplitudes are therefore Rayleigh distributed, and we appeal to the central limit theorem to justify a Gaussian model of their average \citep{Wilensky_2019}. Since the Rayleigh distribution is a 1-parameter distribution, there is a relationship between the resulting mean and variance of the resulting Gaussian, reflected by the aforementioned constant, and the product $N_\text{bl}N_\text{chan}$, which totals the number of elements that went into the average. $N_\text{bl}$ and $N_\text{chan}$ correspond to the number of baselines\footnote{Roughly 8000 for the MWA.} and number of fine frequency channels averaged together when forming the incoherent noise spectrum (cf. Section~\ref{sec:ssins}). The noise is calculated per-\textit{night}, and we take $\hat{\mu}$ to be the median sample for that night. 

In theory, this relationship should apply per-\textit{sample}, but due to the presence of RFI, it is more robust to just use the median sample for all samples in a given night, which theoretically generates $\sim1\%$ errors in the noise standard deviation of any given sample. Since in our use case the compression from any given night's time series to its Legendre coefficients is roughly $1:30$, we expect this error in the noise scale to be completely negligible.\footnote{In other words, we are supposing that the dominant effect of an error in the noise standard deviation is to adjust Legendre coefficient estimates, which take up the bulk of our parameter space and whose values ultimately affect parameters deeper in the hierarchical model. If the time series were sufficiently long, specifically a compression of $\sqrt{1\%} = 1:10000$, we would expect statistically significant differences in the parameter estimation owing strictly to this difference in the noise model. }

Estimating the noise properties of the data depends strongly on the specific application, and must therefore be given careful thought. In cases where these properties are more difficult to define from first principles, the noise may instead be treated as a model parameter and inferred directly from the data.

\subsection{Anomalies due to RFI}\label{sec:rfi}

\begin{figure*}
    \centering
    \includegraphics[width=1.0\linewidth]{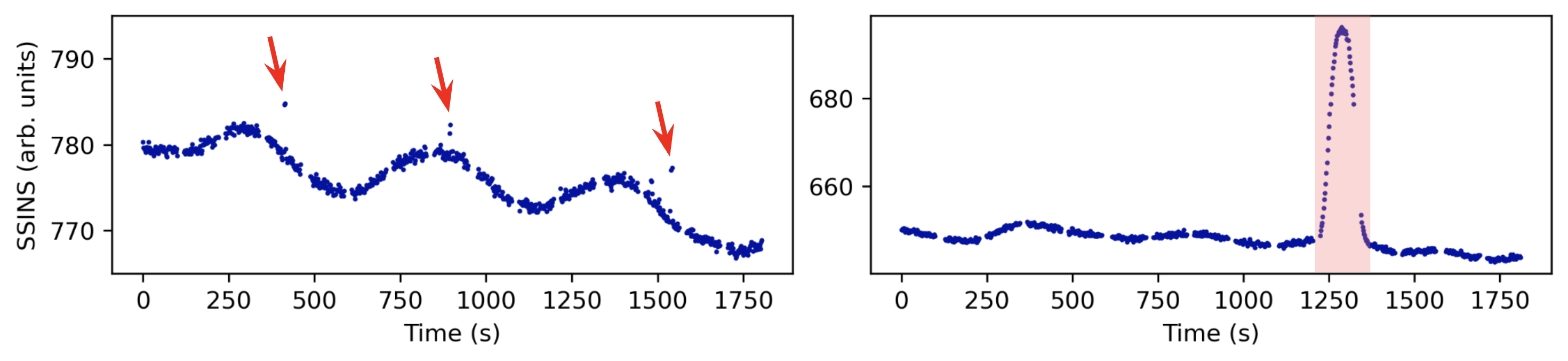}
    \caption{Examples of input time series taken from two different nights, displaying two different classes of RFI events. Left: short-duration, ``blip" anomalies, indicated by the red arrows. Right: long-duration anomaly characteristic of a moving source of RFI, highlighted in pink. Discontinuities along the time axis occur at observation boundaries, roughly every two minutes. The horizontal axis shows the elapsed time, in seconds, since the beginning of each night. Note the difference in scale between both vertical axes.}
    \label{fig:example_rfi_timeseries}
\end{figure*}

RFI originates from a variety of sources and thus leaves a variety of possible signatures in interferometric data. Figure~\ref{fig:example_rfi_timeseries} shows examples of two characteristic signatures of RFI in our input time series. Specifically, the left panel shows examples of short duration ``blip" events, while the right panel shows an example of a longer duration event, with the characteristic rising and falling structure associated with moving sources of RFI.

Despite the diversity of RFI sources and propagation mechanisms, their contamination patterns share certain characteristics. Empirical studies of low-frequency interferometric data have shown that the RFI-dominated tail of visibility amplitude distributions approximately follows a power law, consistent with a large population of weak events and a progressively smaller number of strong events \citep{Offringa_2013}.

However, the distribution of RFI in the SSINS need not preserve the form of the distribution in the raw visibilities, since the data undergo several transformations during the construction of the SSINS data product. We do not attempt to derive the resulting RFI distribution explicitly. Instead, we model the RFI component phenomenologically with a positively skewed Cauchy distribution, which captures the expected concentration of RFI amplitudes while allowing for a heavy tail of rare, exceptionally bright events.

A more complete treatment of RFI would involve modeling the different contamination mechanisms individually. This, however, involves increasing the number of free parameters and requires making additional assumptions. For simplicity, we instead choose to model all RFI with the same skewed Cauchy distribution. As we show in Section~\ref{sec:specific_example}, this distribution captures the statistics of the combined RFI to reasonable accuracy.

\subsection{Probabilistic model}\label{sec:mixture_model}

Here, we introduce the mathematical framework for a mixture model that jointly describes the three components of our input time series.
We start by defining the residual time series for night n as
\begin{equation}
    \mathbf{z}^\mathrm{n}\equiv \mathbf{y}^\mathrm{n} - \mathbf{b}^\mathrm{n},
\end{equation}
where $\mathbf{y}^\mathrm{n}$ is the input time series for night n, and $\mathbf{b}^\mathrm{n}$ is the fit background for that night from Equation \ref{eq:legendre_bg}. For uncontaminated data, the residuals are expected to be consistent with zero within the thermal noise level. We therefore model the clean residual at time index $t$ on night n as
\begin{equation}
p_\text{clean}(z_t^\mathrm{n} \mid \boldsymbol{\theta}) = \mathcal{N}(z_t^\mathrm{n} \mid 0, (\sigma_\text{noise}^\mathrm{n})^2),
\end{equation}
with $\sigma_\mathrm{noise}^\mathrm{n}$ given by Equation \ref{eq:noise}. Here, $\boldsymbol{\theta}$ represents the set of all inferred model parameters to be explicitly defined at the end of this section. As discussed in the previous section, RFI-contaminated residuals are modeled using a skewed Cauchy distribution. Given the general form of the skewed Cauchy probability distribution function:
\begin{equation}
    f_\text{SC}(x \mid \mu,\sigma,a) = \frac{1}{
\sigma\pi
\left[
1+
\frac{|x-\mu|^2}
{\sigma^2\left(1+a\,\operatorname{sgn}(x-\mu)\right)^2}
\right]
},
\end{equation}
we adopt the following parameterization:
\begin{equation}
p_\text{RFI}(z_t^\mathrm{n} \mid \boldsymbol{\theta}) = f_\text{SC}(z_t^\mathrm{n} \mid \mu_\text{RFI}, \sigma_\text{RFI}, a=0.99),
\end{equation}
where $\mu_\mathrm{RFI}$ and $\sigma_\mathrm{RFI}$ correspond to the location and scale parameters, estimated during inference. Under the parameterization adopted here, the skewness parameter satisfies $0 < a < 1$. We set $a = 0.99$, close to its upper bound, to constrain the bulk of the probability mass to positive values while retaining support over the full real line.

Because it is not known a priori whether a given sample is clean or contaminated, we model each sample as arising from either the clean or RFI-contaminated component. This two-component \textit{mixture model} of the residuals allows the class assignment to remain latent while the model parameters are jointly inferred. Specifically, for each residual sample $z_t^\mathrm{n}$,
\begin{equation}\label{eq:mixture_model}
p(z_t^\mathrm{n} | \boldsymbol{\theta}) =
w_\text{clean} \, p_\text{clean}(z_t^\mathrm{n} | \boldsymbol{\theta})
+
w_\text{RFI} \, p_\text{RFI}(z_t^\mathrm{n} | \boldsymbol{\theta}),
\end{equation}
where $w_\text{clean}$ and $w_\text{RFI}$ are the mixture weights satisfying
\begin{equation}\label{eq:mixture_weights}
w_\text{clean} + w_\text{RFI} = 1.
\end{equation}
Assuming that the residual samples are conditionally independent given the model parameters\footnote{RFI contamination often persists in time, such that adjacent time steps are not entirely independent. We discuss possible extensions that account for this temporal dependence in Section~\ref{sec:limitations}.}, the likelihood across all $N$ nights with $T_\mathrm{n}$ time steps each is
\begin{equation}\label{eq}
\mathcal{L}(\boldsymbol{\theta})
=
\prod_{\mathrm{n}=1}^{N}
\prod_{t=1}^{T_\mathrm{n}}
p\left(z_t^\mathrm{n} \mid \boldsymbol{\theta}\right),
\end{equation}
where $\boldsymbol{\theta} = \{ \beta_\ell^\mathrm{n}, \mu_{\beta_\ell}, \sigma_{\beta_\ell}, \mu_\mathrm{RFI}, \sigma_\mathrm{RFI}, w_\mathrm{RFI} \}$ denotes the full set of model parameters.
\begin{figure*}
    \centering
    \includegraphics[width=0.9\linewidth]{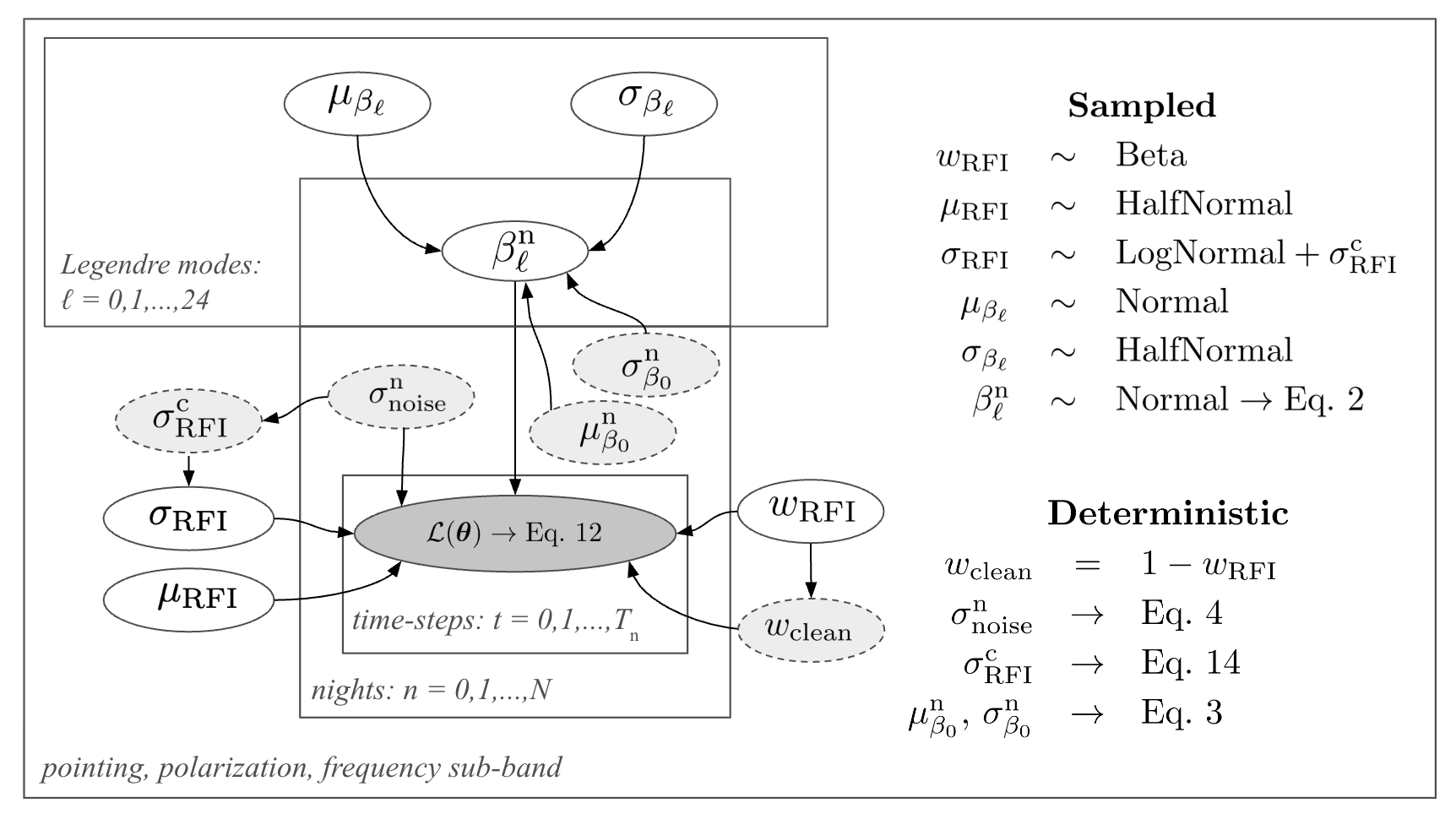}
    \caption{A Bayesian factor graph for the mixture model described in Section~\ref{sec:mixture_model}. Sampled parameters are represented by solid ellipses, whereas dashed ellipses correspond to deterministic parameters.}
    \label{fig:dag}
\end{figure*}
Our complete probabilistic model is summarized in the Bayesian factor graph represented in Figure~\ref{fig:dag}. In this figure, the solid rectangular plates represent repeated dimensions of the input data, with the corresponding dimension and index specified on the bottom left corner of each plate. Reading the graph from the outermost plates inward, the model is applied independently to each combination of pointing, polarization, and frequency sub-band (see Section~\ref{sec:data_compression}). For each such combination, the data consists of $N$ distinct nightly time series, each containing $T_\mathrm{n}$ time steps.

Most model parameters, including the mixing weights and the scale and location parameters of the skewed Cauchy distribution, are shared across nights and time-steps. Other quantities vary along one or more axes of the data. For example, the noise standard deviation introduced in Section~\ref{sec:noise} is estimated separately for each night, while the Legendre coefficients are indexed both by night and Legendre mode, as described in Section~\ref{sec:background}.

We additionally impose a lower cutoff, $\sigma_\text{RFI}^\mathrm{c}$, on the scale parameter of the skewed Cauchy component:
\begin{equation}
\sigma_{\mathrm{RFI}}
\geq
\sigma_{\mathrm{RFI}}^{\mathrm{c}},
\end{equation}
with
\begin{equation}\label{eq:sigma_cutoff}
\sigma_{\mathrm{RFI}}^{\mathrm{c}}
=
2\med
\left(\sigma_{\mathrm{noise}}^\mathrm{n}\right),
\end{equation}
where the median is taken across nights. Without this constraint, the sampled value of $\sigma_{\text{RFI}}$ can approach the characteristic nightly noise scale, causing the clean and RFI-contaminated components to become nearly indistinguishable. We empirically set $\sigma_\mathrm{RFI}^\mathrm{c}$ to twice the median noise standard deviation to preserve the identifiability of the two components and improve numerical stability during inference. As we show in Section~\ref{sec:specific_example}, the sampled posterior estimates for $\sigma_\text{RFI}$ obtained in our analyses remain well above this lower bound, indicating that the cutoff does not constrain the inferred RFI scale in practice.

The priors for $w_\text{RFI}$, $\mu_\text{RFI}$, $\sigma_\text{RFI}$, $\mu_\beta$, and $\sigma_\beta$ are shown in their general forms in Figure~\ref{fig:dag}. The choice of prior for each parameter is motivated primarily by its allowed range. Because $w_\text{RFI}$ is restricted to the interval between 0 and 1, we assign it a Beta prior. The probability density function for the Beta distribution is defined as
\begin{equation}
    f_\mathrm{Beta}(x \mid \alpha, \beta) = \frac{x^{\alpha-1}(1-x)^{\beta-1}}{\mathrm{B}(\alpha, \beta)},
\end{equation}
with
\begin{equation}
    \mathrm{B}(\alpha,\beta) = \frac{\Gamma(\alpha)\Gamma(\beta)}{\Gamma(\alpha+\beta)},
\end{equation}
where $\Gamma$ is the Gamma function, and $\alpha$ and $\beta$ are the shape parameters.

The parameters $\mu_\text{RFI}$ and $\sigma_\beta$ are strictly positive and are therefore assigned half-normal priors, corresponding to the positive half of a zero-centered normal distribution. In contrast, $\mu_\beta$ can adopt any real value, motivating a normal prior. Finally, $\sigma_\text{RFI}$ must be positive and exceed the imposed cutoff $\sigma_\mathrm{RFI}^\mathrm{c}$, so we assign it a shifted log-normal prior. A log-normal random variable is obtained by exponentiating a normally distributed variable and is therefore strictly positive; shifting it by $\sigma_\mathrm{RFI}^\mathrm{c}$ ensures that $\sigma_\text{RFI}>\sigma_\mathrm{RFI}^\mathrm{c}$. Examples of the corresponding hyperparameter choices used in our real-data analyses are provided in Section~\ref{sec:specific_priors}.

\subsection{Classification}\label{sec:classification}

In addition to characterizing the underlying RFI population, our framework allows for the classification of specific samples as either clean or RFI-contaminated. We adopt as a classification scheme the posterior responsibility equation \citep[p.~432]{bishop2006}:
\begin{equation}\label{eq:classification}
\begin{split}
\mathcal{R}_t &= p(\mathrm{RFI} \mid z_t^\mathrm{n}, w_\text{RFI}) \\
&=
\frac{
    w_\mathrm{RFI} \, p_\mathrm{RFI}(z_t^\mathrm{n})
}{
    w_\mathrm{clean} \, p_\mathrm{clean}(z_t^\mathrm{n})
    +
    w_\mathrm{RFI} \, p_\mathrm{RFI}(z_t^\mathrm{n})
}.
\end{split}
\end{equation}
The responsibility $\mathcal{R}_t \in [0,1]$ quantifies the probability that a given sample belongs to the RFI component under the inferred mixture model. Values near zero indicate strong preference for the clean component, while values near unity instead indicate preference for RFI contamination. Intermediate values correspond to samples for which the two components cannot be distinguished with high confidence.

Rather than immediately reducing $\mathcal{R}_t$ to a binary flag, we retain it as a \textit{soft} classification label. This preserves information about classification confidence and allows ambiguous samples to be distinguished from those that are clearly clean or contaminated. A conventional binary mask can be recovered at any stage by applying a threshold to $\mathcal{R}_t$.

Because the mixture parameters are inferred probabilistically, the responsibility itself inherits posterior uncertainty. We therefore evaluate Equation \ref{eq:classification} across posterior samples, obtaining a posterior distribution of $\mathcal{R}_t$ at each time step. This allows both the classification probability and its associated uncertainty to be propagated into subsequent analyses.

\section{Data pre-processing}\label{sec:preprocessing}

The raw input data considered in our case study consist of complex interferometric visibilities organized into a series of two-minute observations. To transform these data into one-dimensional time series, several pre-processing steps are applied.

First, each two-minute observation is processed via \textsc{SSINS}, as described in Section~\ref{sec:ssins}, which collapses the baseline axis. In Section~\ref{sec:data_compression}, we further collapse the polarization and frequency axes to produce one-dimensional time series. In Section~\ref{sec:qa}, we introduce a quality assurance metric developed to identify pathological observations before they can propagate through and bias the analysis. Finally, Section~\ref{sec:model-tailoring} motivates our choice of the number of Legendre modes used for modeling the background fluctuations.

\subsection{SSINS}
\label{sec:ssins}

The Sky-Subtracted Incoherent Noise Spectra (SSINS) is a data product with an accompanying software pipeline\footnote{\url{https://github.com/mwilensky768/SSINS}} for identifying RFI in radio interferometer data, developed by \citet{Wilensky_2019}. This data product is formed in two main steps:
\begin{enumerate}
    \item Sky Subtraction: The difference between time-adjacent complex visibility data from each baseline at each frequency and for each polarization pair. This acts as a high-pass filter that subtracts out the slowly varying astrophysical signal but leaves thermal noise fluctuations and most sources of RFI \citep{Kolopanis2023, Kunicki_Pober_2024}.

    \item Incoherent Averaging: The amplitudes of these finite differences are averaged down the baseline axis, resulting in one dynamic spectrum per polarization pair. Usually auto-baselines (same antenna) are excluded. We refer to these spectra as \textit{incoherent noise spectra} or just the SSINS. While not all baselines may see a given RFI source at the same amplitude, this step empirically tends to boost the signal-to-noise ratio of RFI sources, allowing for deeper RFI excision than methods that operate one baseline at a time (e.g. \citealt{Offringa2015, Wilensky_2019, Barry2019, Wilensky_2023, Kunicki_Pober_2024}, though see \citealt{Offringa2023} for a counterexample).
\end{enumerate}

The software for making SSINS is co-packaged with a customizable flagging algorithm developed on MWA data. In brief, the user first provides a dictionary of frequency sub-bands occupied by known RFI contaminants, such as 7 MHz bands corresponding to digital TV allocations in Australia, which are shown in Table \ref{tab:freq_subbands}. Then, an iterative match filter is applied alternating between mean estimation via a time average (to estimate the background) and masking statistically significant matches for the filter. A match for the filter can be formulated as a Bayesian maximum \textit{a posteriori} decision rule between two models for the data: one where the data is noise-like, and another where the data has been biased by a constant value implicitly determined by the user-set significance threshold. This is equivalent to modeling the RFI amplitudes as a delta-function distribution i.e. constant amplitude. While this model is surprisingly effective for generating flags given how radically incorrect it is,\footnote{Note, however, decision rules and model comparisons do not map one-to-one i.e. there may be some more reasonable model that implicitly produces the same decision rule.} the results from flagging give no immediate way of understanding the identified RFI without extensive post-processing. Explicitly modeling the RFI amplitude distribution in SSINS is a major improvement that we develop in this work that allows the user to more effectively characterize and treat RFI in their data, including making more accurate flags if they wish. 

As discussed in Section~\ref{sec:noise}, the noise statistics of the SSINS are theoretically tractable as they are Gaussian and negligibly heteroskedastic. The co-packaged flagging software is usually applied to SSINS in small chunks of time (e.g. 2 minutes), in order to help avoid slowly varying drifts in the SSINS that arise from overall gain fluctuations from the refrigeration cycle. These drifts can appear as broadband RFI in the SSINS, causing false positives. Explicitly modeling these slowly varying fluctuations is a major improvement resulting from this work that significantly reduces the false positive rate.

We compare results from this method with \textsc{SSINS} flags in Section~\ref{sec:model_checking}. The exact \textsc{SSINS} flagging settings applied to this data are listed by \citet{Wilensky_2023}. In short, data were flagged in 2-minute intervals, and RFI shapes were made for each sub-band we enumerate in this work (Section~\ref{sec:data_compression}), with a significance threshold of 5$\sigma$ for all sub-bands except for ``broadband streaks" (i.e. averaging the entire band), which used a significance threshold of 10$\sigma$. Additionally, we do not explore narrowband RFI characterization in this work, though it is standard to search for narrowband RFI in the \textsc{SSINS} pipeline. Since narrowband RFI is the most ubiquitous RFI in radio astronomy, we remark that all of the techniques we develop work at any frequency resolution, and so narrowband RFI characterization can be accomplished with these techniques by simply applying them on a channel-by-channel basis. 

\subsection{Data compression}
\label{sec:data_compression}

After processing via \textsc{SSINS}, our input data remains a function of time, frequency, and polarization. We further compress our input space by selecting five frequency sub-bands that are commonly contaminated by RFI, and averaging across each of these sub-bands separately. We select the frequency channels corresponding to Australian digital television broadcast bands due to the high occurrence rate of RFI in these specific bands in MWA data. These are listed in Table \ref{tab:freq_subbands}, and highlighted for visual reference in Figure~\ref{fig:waterfall}.

\begin{table}
    \centering
    \begin{tabular}{|c|c|}
        \hline
        Name & Frequency range (MHz) \\
        \hline
        \texttt{subTV} & 167.1 -- 174   \\
        \texttt{TV6}   & 174 -- 181     \\
        \texttt{TV7}   & 181 -- 188     \\
        \texttt{TV8}   & 188 -- 195     \\
        \texttt{TV9}   & 195 -- 197.7   \\
        \hline
    \end{tabular}
    \caption{Frequency sub-bands corresponding to Australian digital television allocations \citep{dtv_allocations}. The lower boundary of the \texttt{subTV} band, 167.055 MHz, and the upper boundary of the \texttt{TV9} band, 197.735 MHz, are determined by the frequency coverage of the observations used in this work rather than by digital television allocation boundaries. Note that the \texttt{subTV} band does not correspond to an official digital television allocation; instead, it denotes frequencies below the allocated television bands.}
    \label{tab:freq_subbands}
\end{table}

This produces five time series per polarization. Averaging along several frequency channels not only helps reduce ultimate input size, but it also reduces noise in the resulting time series. We further restrict our input space by considering only the XX and YY polarization modes, corresponding, respectively, to east-west and north-south orientations.

As discussed in Section~\ref{sec:background}, the temperature-dependent background fluctuations are expected to vary smoothly within a night. Individual two-minute observations are thus grouped by night, where each night's background fluctuations are modeled independently.

Furthermore, we expect the statistical properties of our time series to depend on the specific \textit{pointing} considered. Here, a pointing denotes a particular electronic beam-steering configuration of the MWA. The MWA consists of antenna tiles, each comprising 16 dual-polarization dipoles arranged in a $4\times4$ grid. The analog beamformer applies relative delays to the dipoles within each tile, directing the tile primary beam toward a selected region of the sky \citep{mwa1}. Because the beamformer delays can take only discrete values, the MWA observes using a discrete set of pointing configurations rather than continuously variable pointing directions. Different pointings correspond to different sets of beamformer delays, and hence to different primary-beam responses and regions of the sky.

We therefore group the input nightly time series by pointing. For the data considered here, five distinct pointings are available. These are labeled from $-2$ to $+2$ inclusively, with pointing 0 corresponding to zenith. These five pointings form a subset of the MWA's ``east-west" pointings, which themselves constitute only a subset of all available instrumental pointing configurations. The altitude and azimuth associated with each pointing are listed in Table \ref{tab:pointings}.

\begin{table}
    \centering
    \begin{tabular}{|c|c|c|}
        \hline
        Pointing & Altitude ($^\circ$) & Azimuth ($^\circ$) \\
        \hline
        -2 & 76.28 & 90 \\
        -1 & 83.19 & 90 \\
        0 & 90 & 0 \\
        +1 & 83.19 & 270 \\
        +2 & 76.28 & 270 \\
        \hline
    \end{tabular}
    \caption{List of MWA pointings considered in this work. The pointing directions correspond to a subset of the available MWA beamformer grid positions; altitude and azimuth values are taken from the MWA pointing-grid documentation \citep{mwa_sweetspots}. Note that pointing 0 corresponds to zenith, and that azimuth is measured clockwise from north.}
    \label{tab:pointings}
\end{table}

We take a conservative approach and apply our probabilistic model independently to each combination of pointing, polarization, and frequency sub-band, without assuming identical RFI statistics across these combinations. This allows for potential differences in the contamination without imposing assumptions about their physical origin. The inference is performed jointly across all nights associated with each combination. Given five pointings, two polarization modes, and five frequency sub-bands, this results in 50 independent model fits. This structure is illustrated in Figure~\ref{fig:dag}. The outermost rectangular plate is defined for a specific pointing, polarization, and frequency sub-band, as indicated by the annotation in the lower-left corner of the plate. Some parameters, such as the RFI location and scale parameters, or the mixture weights, are jointly inferred across all nights. Other parameters, such as the Legendre coefficients used to model the smooth background, are inferred independently for each night.

The choice of axes along which to partition the data is dependent on the specific use case. For the MWA observations considered here, pointing, polarization, and frequency sub-band provide natural divisions because each is associated with a distinct source of variation in RFI contamination. Other instruments, observing strategies, or RFI environments may motivate a different partitioning of the data.

\subsection{Quality assurance}\label{sec:qa}

To prevent outlier data from biasing our ultimate modeling framework, we run a simple quality assurance test on each night: if more than 1/3 of that night's samples are flagged by the default \textsc{SSINS} flagging algorithm (described in Section~\ref{sec:ssins}), the whole night is marked as contaminated and removed from the input space.

One should note that the built-in \textsc{SSINS} flagger is sensitive to fluctuations in the thermal background and requires careful tuning to reduce the rate of false positives. Poorly calibrated settings can cause the criterion described above to reject nights containing substantial amounts of otherwise usable data. Developing a more robust quality assurance heuristic that does not require prior flagging, potentially based on diagnostics derived directly from the data, is left for future investigation.

\subsection{Model tailoring}\label{sec:model-tailoring}

In order to avoid overfitting, the Legendre polynomial series are truncated at $L = 24$ total terms. To arrive at this upper bound, we determine best-fit Legendre coefficients by looking at a subset of hand-picked time series that are qualitatively clean. For this analysis, we depart from the mixture model and directly calculate linear best fits by Legendre polynomials to these time series. We then compare the obtained coefficients to the diagonal terms in the posterior coefficient covariance for each fitted time series, to obtain an upper truncation limit for the Legendre basis and some intuition about the strength of the Legendre coefficients as a function of \(\ell\). The propagated uncertainty in best fit coefficients is denoted as $\boldsymbol \Sigma_p$, arising from the thermal noise in our data. Legendre coefficient covariance is given by
\begin{equation}
    \boldsymbol{\Sigma}_p = \left(\mathbf{D}^{\text{T}}\mathbf N^{-1} \mathbf D\right)^{-1},
\end{equation}
where $\mathbf D$ is the Legendre design matrix and $\mathbf N$ is the noise covariance matrix, given as the diagonal matrix of the thermal noise $\sigma_\mathrm{noise}^\mathrm{n}$ for each time step. Though each noise matrix is computed independently per night, the order of the error is roughly the same per night.

\begin{figure}[h]
    \centering
    \includegraphics[width=\linewidth]{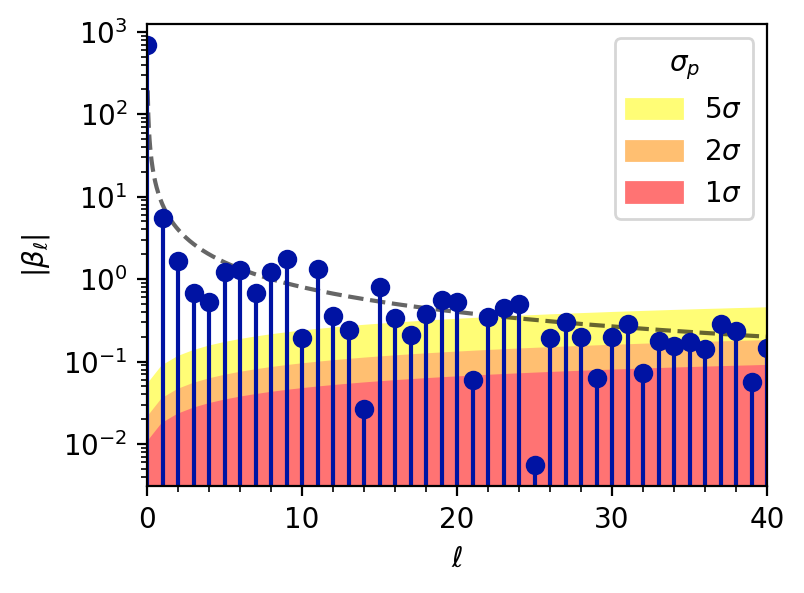}
    \caption{Legendre best-fit coefficients $|\beta_\ell|$ for $0 \leq \ell \leq 40$ for the night of 2014-08-23 in pointing 0 across the TV7 frequency band for the XX polarization mode (blue). Parameter significance $\sigma_p$, the diagonal elements of $\boldsymbol\Sigma_p$, is plotted in comparison for thresholds $1\sigma$ (red), $2\sigma$ (orange), and $5\sigma$ (yellow). The curve $8/\ell$ is plotted for visual reference (dotted, black). Best-fit Legendre coefficients cross underneath $5\sigma$ in parameter significance at $\ell \approx 24$.}
    \label{fig:sigma_p_leg}
\end{figure}

In Figure~\ref{fig:sigma_p_leg}, best-fit coefficients for a single pointing and night are compared to the posterior coefficient uncertainties for $N = 40$ terms. Coefficient value is found to decrease with term number, while parameter variance $\sigma_p^2$ increases. In particular, best-fit Legendre coefficients begin to dip below $5\sigma_p$ past roughly $\ell \approx 24$ terms in the time series we study. Coefficient values too close to coefficient uncertainty cannot be distinguished between thermal noise contribution and any meaningful morphology in the smooth background. We thus choose $N = 24$ terms as a natural truncation point for the Legendre basis.

\section{Implementation}\label{sec:implementation}

We implement our joint modeling framework using \texttt{numpyro}\footnote{\url{https://num.pyro.ai/en/stable/}} \citep{numpyro}. We run Hamiltonian Monte Carlo using the NUTS sampling strategy \citep{hoffman2014} to produce samples from the joint posterior distribution. Sampled parameters include the Legendre coefficients (per-night), the hierarchical Legendre modeling parameters $\mu_{\beta_\ell}$ and $\sigma_{\beta_\ell}$, the location and scale parameters for the skewed Cauchy distribution, and the RFI mixing weight. We specify prior distributions for each of these parameters, as detailed in Section~\ref{sec:specific_priors}.

Each input time series is sampled using 48 independent chains a total of 9000 times each. The first 6000 samples serve as warm-up and are ultimately discarded.

In a small number of cases, typically one or two of the 48 chains, sampling becomes trapped in a region where the inferred value of $\sigma_\text{RFI}$ approaches $\sigma_\text{noise}$, despite the lower cutoff $\sigma_\mathrm{RFI}^\mathrm{c}$. In this limit, the clean and RFI components become difficult for the model to distinguish. The corresponding latent-state assignments therefore become weakly identified, producing a poorly constrained region of the posterior in which a chain may remain for an extended period and mix poorly with the others.

Rather than increasing $\sigma_\mathrm{RFI}^\mathrm{c}$, we identify such chains using a modified version of the Gelman-Rubin convergence diagnostic, which compares the variability within chains to the variability across chains \citep{gelman_rubin}. For each chain, we compare the standard deviation of its post warm-up samples with the standard deviation of the samples pooled across all 48 chains. If the within-chain standard deviation is more than a factor of 1000 smaller than the pooled standard deviation, we classify the chain as stuck and exclude it from subsequent analysis.

\subsection{Compute requirements}

The mixture model is fit independently for each combination of pointing, frequency sub-band, and polarization, resulting in 50 independent fits in our implementation. As a representative example of the computational requirements, a single fit with 48 NUTS chains (6000 warmup and 3000 sampling iterations each) run on 12 CPU cores of a dual-socket AMD EPYC 9684X node reaches a peak memory usage of approximately 9 GB of RAM. Wall-clock time varies from fit-to-fit but typically ranges from 10 to 40 minutes, depending on the efficiency with which chains explore the posterior. Because each pointing/frequency/polarization combination is independent, these fits can be executed in parallel. The main computational limitation of the mixture-model stage is therefore the number of CPU cores available for parallel execution.

In our case study, however, the dominant computational bottleneck occurs earlier in the pipeline, during the conversion of the raw interferometric visibilities into the one-dimensional time-series representation described in Section~\ref{sec:data_compression}. Each 2-minute MWA observation must be loaded into memory, processed with \textsc{SSINS}, and compressed into the required time-series format. This stage can also be parallelized across observing nights, each of which typically contains approximately 40 observations, but the processing of a full night can still require several hours.

\subsection{Code availability}

The code used to perform the analysis presented in this work (version 1.0.0) is publicly available at \url{https://github.com/jade-ducharme/bayesian-rfi/releases/tag/v1.0.0}.

\section{The 2014 MWA observing season as a worked example}\label{sec:real_analysis}

We use the same data as \citet{Wilensky_2023}, with all the same pre-processing, summarized in Section~\ref{sec:preprocessing}. These observations target the EoR0 field (RA = 0 h, Dec = -27 deg) and come from the 2014 MWA Phase I EoR highband data (167.7--197.8 MHz). Before quality cuts, the data span July 17 to September 14, 2014, and represent a subset of the full EoR observing season. \citet{Wilensky_2023} performed a \textsc{SSINS}-based RFI analysis of 3168 two-minute observations (98.56 hours). We make use of the uncalibrated SSINS and the flags generated from them in that analysis. According to the quality assurance metric described in Section~\ref{sec:qa}, approximately 2\% of nights were rejected and thus not considered here.

\subsection{Prior hyperparameters}\label{sec:specific_priors}

The prior distributions for all sampled parameters are presented in their general forms in Figure~\ref{fig:dag}. We now detail the specific hyperparameters chosen, although we note that these choices are highly dependent on our specific use case, and will require careful tuning for other applications. Specifically, we adopt
\begin{equation}
    \begin{split}
        w_\mathrm{RFI} &\sim \mathrm{Beta}(2, 10), \\
        \mu_\mathrm{RFI} &\sim \mathrm{HalfNormal}(0.25), \\
        \sigma_\mathrm{RFI} &\sim \mathrm{LogNormal}(0, 0.5) + \sigma_\mathrm{RFI}^\mathrm{c}, \\
        \mu_{\beta_\ell} &\sim \mathrm{Normal}(0, 1), \\
        \sigma_{\beta_\ell} &\sim \mathrm{HalfNormal}(1). \\
    \end{split}
\end{equation}
These priors are intended to be weakly informative. They constrain the parameters to physically meaningful regions and mildly favor expected values for our data, while remaining broad enough for the posterior to be driven primarily by the observations.

In particular, the prior on $w_\mathrm{RFI}$ favors a relatively small contaminated fraction without excluding nights with higher RFI occupancy. The half-normal prior on $\mu_\mathrm{RFI}$ enforces a non-negative RFI amplitude, while the shifted log-normal prior on $\sigma_\mathrm{RFI}$ ensures that the RFI component is distinguishable from the clean component. Finally, the priors on $\mu_{\beta_\ell}$ and $\sigma_{\beta_\ell}$ regularize the Legendre coefficients around zero while allowing significant variation between modes.

\subsection{Representative examples}\label{sec:specific_example}

The mixture model developed in this work operates independently on different combinations of pointing, frequency sub-band, and polarization. While Section~\ref{sec:summary_statistics} demonstrates how to synthesize all these various outputs into season-level conclusions, we focus on representative examples for a few specific combinations here. 

\begin{figure*}
    \centering
    \includegraphics[width=1.0\linewidth]{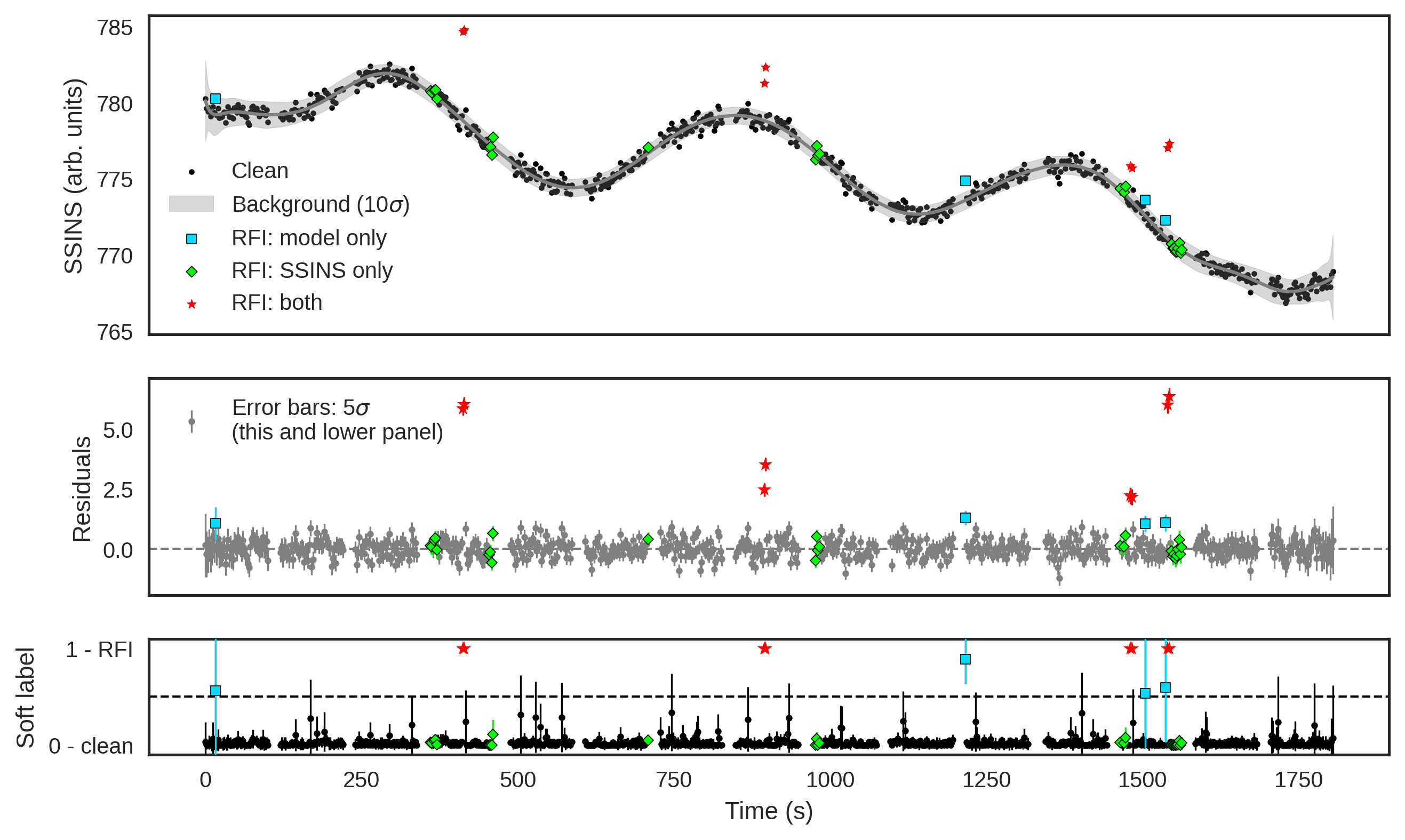}
    \caption{Model output for a single night, for MWA pointing 0. Specifically, this corresponds to the night of 2014-07-25, with observations taken from UTC 19:44:32 to UTC 20:12:56, before and after which the array was electronically steered to a different pointing. The data have been compressed according to Section~\ref{sec:data_compression}, for the TV7 frequency band and the XX polarization mode. Top panel: data points, color-coded by model classification, with black corresponding to clean data, blue corresponding to data labeled as RFI by the mixture model, green to data labeled as RFI by \textsc{SSINS}, and red to data labeled as RFI by both methods. The grey shaded region shows the 10$\sigma$ posterior credibility interval for the Legendre background fit. Middle panel: Residuals taken by subtracting the data with the posterior background, along with 5$\sigma$ error bars. Bottom panel: Soft labels obtained via Equation \ref{eq:classification}, along with 5$\sigma$ error bars. We discuss possible explanations for the discrepancies between the two classification methods in Section~\ref{sec:specific_example}.}
    \label{fig:single_night}
\end{figure*}

For instance, consider the night of 2014-07-25, for MWA pointing 0, averaged across the TV7 frequency band for the XX polarization mode. Data from this night, after processing via Section~\ref{sec:data_compression}, are presented in the top panel of Figure~\ref{fig:single_night}. In gray, we overlay the 10$\sigma$ posterior credibility interval for the smooth background, modeled using a sequence of Legendre polynomials. Qualitatively, we observe a close agreement between the background fit and the data. We note that the uncertainty in the background fit increases near the beginning and end of the time series, where the Legendre polynomial fit is less constrained. Alternative polynomial bases, such as Chebyshev polynomials, may offer different behavior near the boundaries and could be investigated as a possible improvement.

The middle panel shows the residuals obtained after subtracting the background trend from the data, while the bottom panel shows the soft labels obtained via the posterior responsibility from Equation \ref{eq:classification}. The data are furthermore color-coded based on model classification. For the mixture model, we obtain binary classifications by thresholding the posterior mean responsibility at 0.5. Samples with a mean responsibility above 0.5 are classified as RFI, while those below 0.5 are classified as clean. Points classified as RFI only by the mixture model are shown in blue, those classified only by \textsc{SSINS} are shown in green, and those classified by both methods are shown in red.

The \textsc{SSINS} classifications are derived from the flags produced according to Section~\ref{sec:ssins}, restricted to the same TV7, pointing 0, and XX polarization combination. To reduce the \textsc{SSINS} flags to a single classification per time sample, we collapse the frequency axis using a logical OR: a time sample is classified as contaminated if at least one frequency channel within the TV7 band is flagged. This preserves sensitivity to narrowband contamination while producing a binary time series we can directly compare to the mixture model classifications.

We note that \textsc{SSINS} and the mixture model generally agree on clear outliers, while they may disagree on data points lying closer to the fitted background. In particular, \textsc{SSINS} seems to exhibit a relatively high rate of detections towards the edges of observations, particularly when those observations have strong gradients. We believe these are false positives owing to the assumption of a constant background when \textsc{SSINS} was run on this data. These types of false positive detections are a known feature of \textsc{SSINS}, and the option to fit a polynomial to each 2-minute chunk does exist in the \textsc{SSINS} software, rather than assume a constant background. However, the relatively small time window that \textsc{SSINS} is often applied on with MWA data can lead to underflagging even with this amenity, since the per-observation fit can sometimes fit out RFI \citep{Star2024}. The more global analysis with explicit mixture modeling that we present in this work largely overcomes the false positive issue\footnote{One might wonder whether this improvement arises primarily from the background modeling or the probabilistic classification. Applying the default SSINS flagger after subtracting the posterior median Legendre background yields qualitatively similar results, although the mixture model also identifies some weaker RFI events missed by SSINS. However, we caution that since the background was jointly inferred, such comparisons do not completely isolate these individual contributions.}. We expect the overfitting issue is also largely resolved, though see Section~\ref{sec:model_ambiguities}. We also note that RFI overfitting mainly presented with narrowband RFI in previous analyses, which we did not investigate in this work. 

\begin{figure*}
    \centering
    \includegraphics[width=1.0\linewidth]{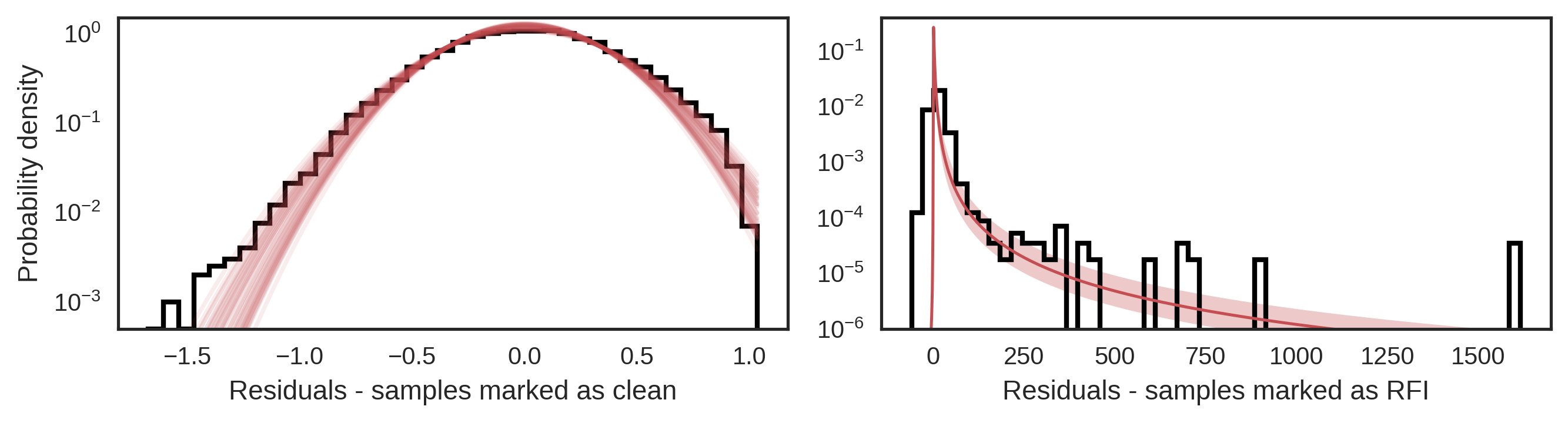}
    \caption{Left: histogram of residuals marked as clean by the mixture model. Overlaid in red are the zero-centered normal distributions with standard deviations corresponding to the calculated values of the noise standard deviation (Eq. \ref{eq:noise}) for all nights in this set. Right: histogram of residuals marked as RFI by the mixture model. Overlaid in red is the $10 \sigma$ region of the expected counts under the skewed Cauchy distribution, computed from the sampled $\mu_\mathrm{RFI}$ and $\sigma_\mathrm{RFI}$ values. These data correspond to all nights for MWA pointing 0, averaged across the TV7 frequency band for the XX polarization.}
    \label{fig:rfi_residuals}
\end{figure*}

The example in Figure~\ref{fig:single_night} shows a single night, while inference is performed jointly across \textit{all} nights for each pointing/frequency/polarization combination. The combination considered here contains 42 nights in total. Rather than analyzing the output plots for each of these nights, we can draw some overall conclusions about model performance by considering the residual distributions of all data points classified as clean or RFI-contaminated based on the hard labels. These are shown in Figure~\ref{fig:rfi_residuals}. Here, the left panel shows the histogram of residuals marked as clean by the mixture model. Zero-centered normal distributions with standard deviations given by the theoretically calculated noise level from Equation \ref{eq:noise} for each night are overlaid in red. We note the good qualitative agreement between the residuals and the expected normal distributions, which supports both the theoretical noise estimates from Section~\ref{sec:noise} and the accuracy of the background model.

Meanwhile, the right panel shows the residuals for the data points labeled as RFI by the mixture model. We overlay a red shaded region corresponding to $\pm 10$ standard deviations of the expected counts under the skewed Cauchy distribution, computed from the sampled $\mu_\mathrm{RFI}$ and $\sigma_\mathrm{RFI}$ values. While the model captures the broad structure of the RFI distribution, including the concentration of samples at lower amplitudes and less frequent brighter events, the residuals exhibit additional structure that is not fully described by the simple model adopted here. More sophisticated RFI models would be required to capture this complexity in greater detail, which we leave for future investigation.

\begin{figure}
    \centering
    \includegraphics[width=1.0\linewidth]{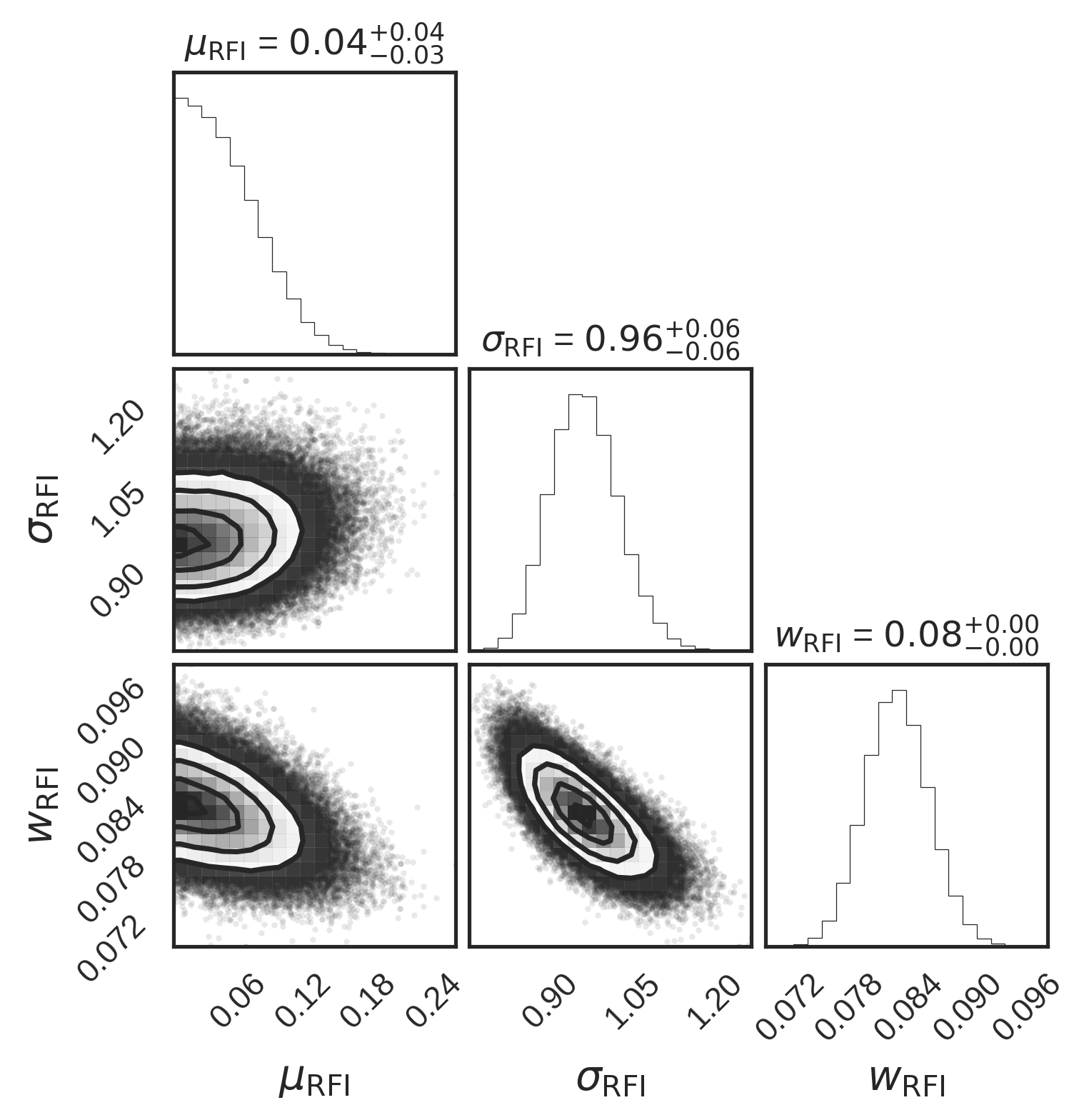}
    \caption{Marginalized 1D and 2D posterior probability distributions for $\mu_\mathrm{RFI}$, $\sigma_\mathrm{RFI}$, and $w_\mathrm{RFI}$, inferred from pointing 0 in the TV7 frequency sub-band and XX polarization. The posteriors are well constrained and seemingly unimodal.}
    \label{fig:corner_plot}
\end{figure}

As an additional assessment of the quality of the Bayesian inference, we examine the marginalized posterior distributions and pairwise parameter correlations using corner plots. Figure~\ref{fig:corner_plot} shows an example for the same pointing, frequency sub-band, and polarization combination. For clarity, we display three representative parameters: the location and scale of the RFI component, $\mu_\mathrm{RFI}$ and $\sigma_\mathrm{RFI}$, and the inferred RFI occupancy, $w_\mathrm{RFI}$. The corresponding posterior distributions are well constrained and seemingly unimodal, while the joint distributions show no evidence of unexpected degeneracies or structures that would suggest poor convergence or sampling failure.

We note that the bulk of the sampled values for $\sigma_\mathrm{RFI} = 0.96 \pm 0.06$ for this particular combination remains well above the calculated value of $\sigma_\mathrm{RFI}^\mathrm{c} = 0.67$, obtained via Equation \ref{eq:sigma_cutoff}, providing additional support for the use of this criterion. We also note that the posterior for $\mu_\mathrm{RFI}$ features a sharp cutoff at 0 due to the imposed positivity constraint, which reflects our assumption that RFI positively biases the SSINS.

As part of our standard convergence checks, we also monitor the Gelman--Rubin statistic, $\hat{R}$, and the effective sample size for all sampled parameters. In nearly all cases, $\hat{R}$ remains below 1.01, while the effective sample sizes are well into the thousands, further supporting good convergence of the sampled posterior distributions.

\subsubsection{Classification ambiguities}\label{sec:model_ambiguities}

\begin{figure*}
    \centering
    \includegraphics[width=1.0\linewidth]{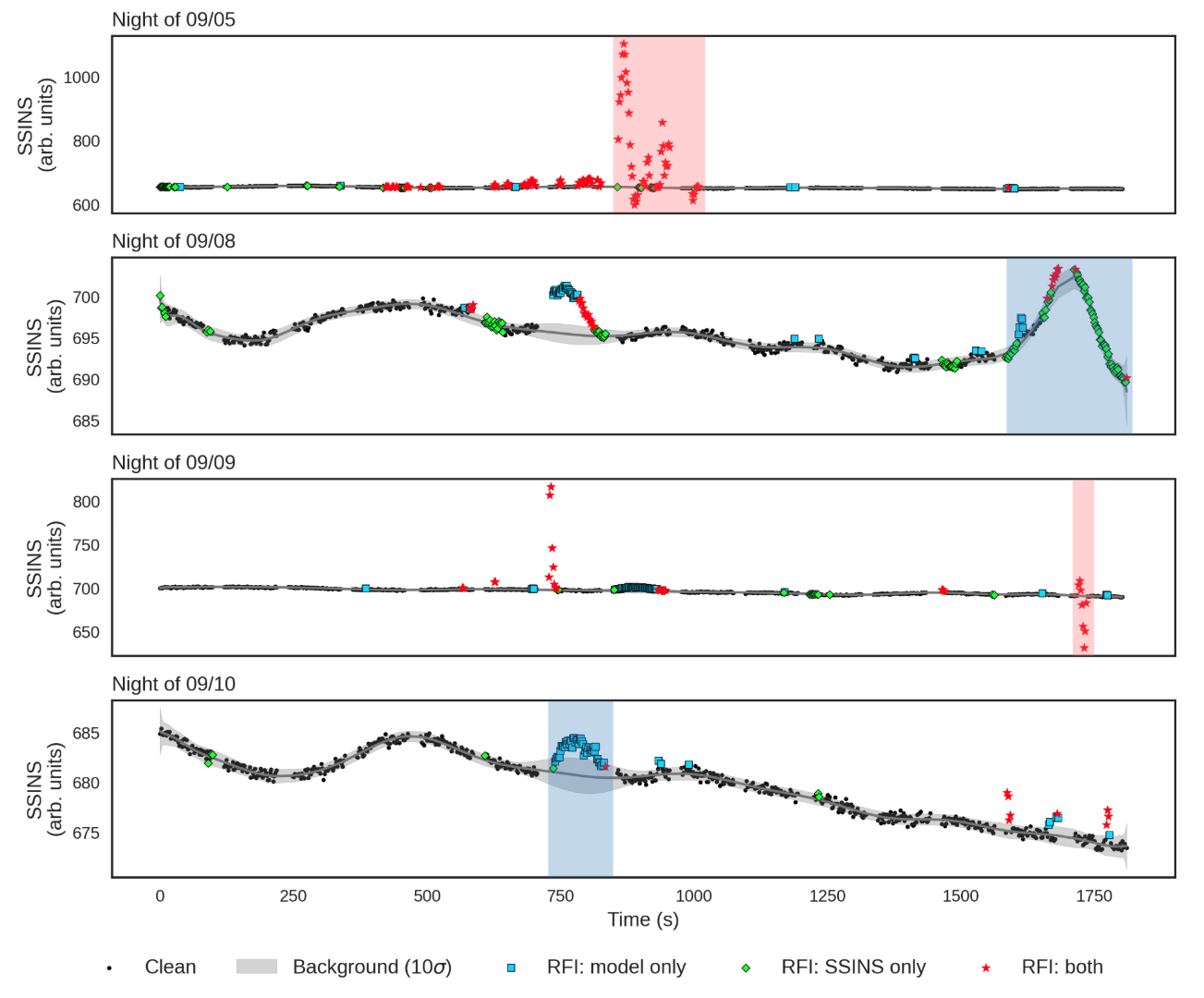}
    \caption{Time series from four nights labeled in MM/DD format illustrating representative instrumental anomalies and classification ambiguities encountered by the mixture model. Instrumental failure modes are highlighted in pink, while ambiguous classifications are highlighted in blue. All examples correspond to pointing 0, TV7 frequency band, and XX polarization. See Section~\ref{sec:model_ambiguities} for further discussion.}
    \label{fig:nights_bad_examples}
\end{figure*}

Although the mixture model developed here generally performs well on the 2014 MWA season data as a whole, with satisfactory convergence according to standard diagnostics such as $\hat{R}$ and effective sample size and with well-behaved posterior distributions in most cases, several limitations should be kept in mind when interpreting its classifications. Representative examples are shown in Figure~\ref{fig:nights_bad_examples}.

The first limitation concerns the interpretation of samples assigned to the contaminated component. As illustrated in the first and third rows of Figure~\ref{fig:nights_bad_examples}, occasional instrumental failures lead to a negative bias in the data, producing sharp negative residuals relative to the inferred background. Although the skewed Cauchy component was introduced primarily to model the strong positive tail characteristic of RFI contamination, the mixture model nevertheless assigns these negative outliers to the contaminated state. Despite lying on the less probable side of the skewed Cauchy distribution, sufficiently large negative residuals are still more likely under this distribution than under the Gaussian distribution describing clean data. The model therefore correctly identifies these samples as inconsistent with the clean state; in other words, ``anomalous".

In these examples, however, the anomalies are instrumental rather than RFI-related. For our application, flagging these outliers is desirable. Data affected by instrumental failures should also be identified and ultimately discarded. More generally, however, applications of this type of mixture model should distinguish carefully between sources of anomalous behavior and consider whether they should all be grouped into a single anomalous class. We include these examples to make this limitation explicit and to illustrate that flagged samples should not automatically be interpreted as RFI according to the framework presented here.

A second limitation arises from the separation of smooth background structure from RFI contamination. In the second row of Figure~\ref{fig:nights_bad_examples}, a relatively smooth and localized excess is absorbed into the Legendre polynomial background instead of being identified as contamination, despite being flagged by \textsc{SSINS}. This discrepancy is inherently ambiguous, as the feature may represent either an unusually sharp component of the background, as favored by the mixture model, or an unusually smooth RFI event, as suggested by the \textsc{SSINS} flags. The converse case appears in the bottom row of Figure~\ref{fig:nights_bad_examples}, where relatively sharp fluctuations are not captured by the Legendre background modeling and are instead assigned to the RFI component, despite not being flagged by \textsc{SSINS}.

Such disagreements are not altogether unexpected, since different RFI-detection methods process data differently and need not produce identical classifications. Rather than treating these cases as evidence that one method is necessarily incorrect, they highlight regions of the data that warrant closer inspection. An advantage of the framework presented here is that its probabilistic classification and associated uncertainty estimates provide additional information for investigating such ambiguous cases. More broadly, the inferred posterior distributions also enable checks along other axes of the model; for example, nights with discrepant classifications can be examined for unusually large or otherwise atypical Legendre coefficients relative to the population-level distribution.

The balance between absorbing smooth RFI into the background and misclassifying sharp background fluctuations as RFI is partly controlled by the flexibility of the background model. Using fewer Legendre coefficients restricts the background fit, making it less likely to absorb smooth RFI events. Conversely, using more coefficients allows the background model to capture more complex intrinsic fluctuations, at the cost of potentially absorbing smooth contamination into the background. The choice therefore reflects a tradeoff between these two forms of misclassification. In this work, we adopt 24 Legendre coefficients, a value selected through careful tuning to balance this tradeoff, as described in Section~\ref{sec:model-tailoring}.

\subsection{Model validation and consistency checks} \label{sec:model_checking}

To ensure that the model behaves as intended, we perform several consistency checks. 

\begin{figure}
    \centering
    \includegraphics[width=1.0\linewidth]{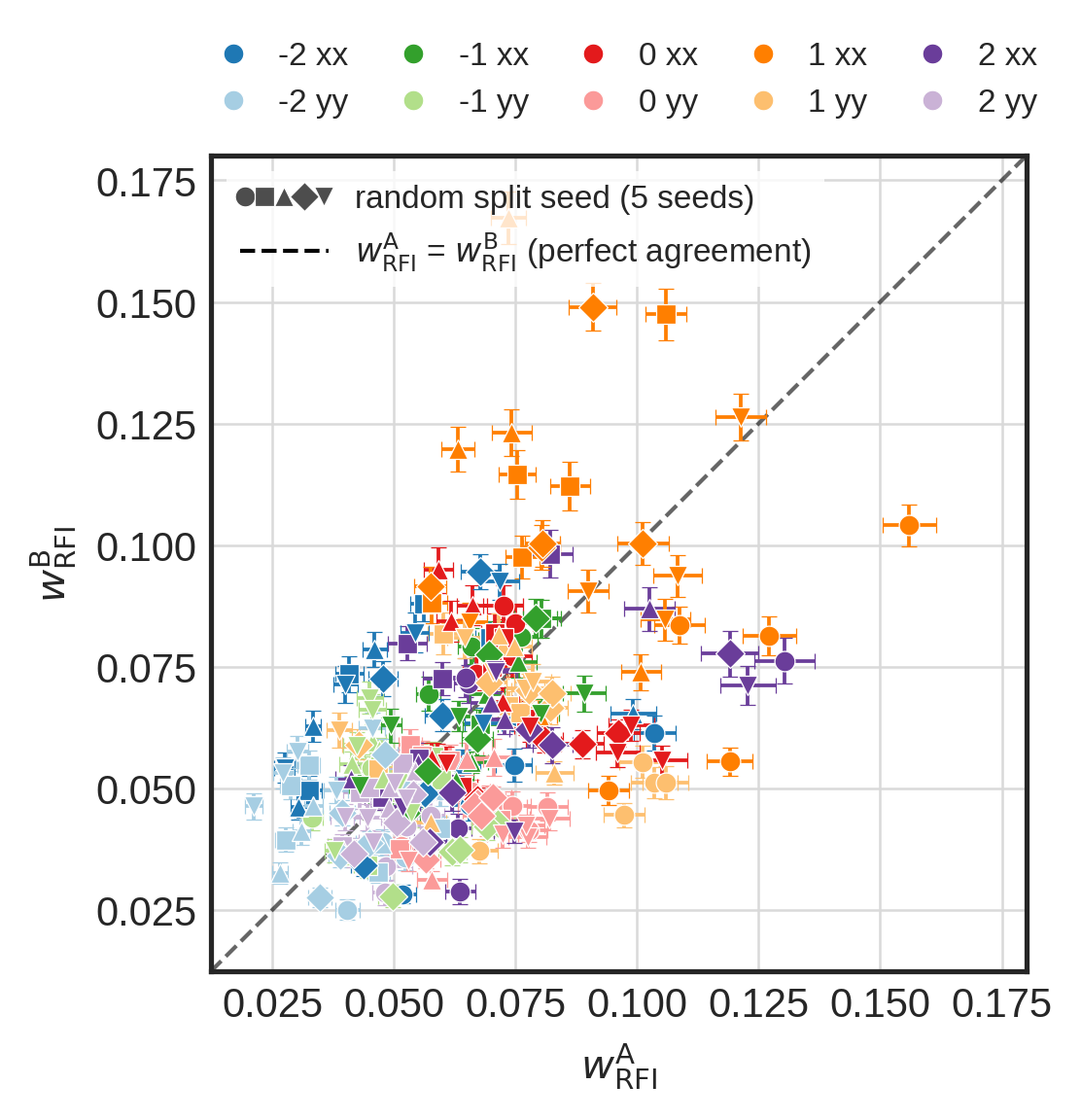}
    \caption{Comparison of the inferred seasonal RFI occupancy between two randomly selected halves of the observing season. The inferred occupancies for the two subsets are labeled $w_\mathrm{RFI}^A$ and $w_\mathrm{RFI}^B$. Colors represent different pointings and polarizations, while markers represent different random seeds.}
    \label{fig:random_split}
\end{figure}

First, we randomly divide the season data into two subsets containing an equal number of nights and run inference on each subset independently. Because the split is random, the inferred season-level parameters should, in principle, be consistent between the two halves up to statistical fluctuations. However, comparison of the inferred parameters, including the seasonal RFI occupancy $w_\mathrm{RFI}$, reveals systematic differences. This is illustrated in Figure~\ref{fig:random_split}, where the inferred $w_\mathrm{RFI}$ values between the two halves (labeled $w_\mathrm{RFI}^A$ and $w_\mathrm{RFI}^B$) are compared for five independent random splits. The parameter is inferred separately for each combination of pointing, polarization, and frequency sub-band. Although the frequency dependence is not explicitly illustrated in the figure, pointing and polarization are distinguished by color, while different random splits are shown using different markers. 

The magnitude of the disagreement is weakly correlated with pointing and polarization. Indeed, we note that the YY polarization tends to cluster at lower values of $w_\mathrm{RFI}$ and correspondingly exhibits a smaller spread between $w_\mathrm{RFI}^A$ and $w_\mathrm{RFI}^B$. Conversely, pointing 1 is associated with larger inferred RFI occupancies and shows larger discrepancies between the two subsets. These seasonal trends are discussed more in depth in Section~\ref{sec:summary_statistics}. Error bars correspond to the 94\% highest-density intervals (HDIs) derived from the posterior distributions.

Interestingly, most inferred $w_\mathrm{RFI}$ values from the two random halves are inconsistent within their posterior uncertainties, regardless of the particular random split. We do not interpret this discrepancy as evidence of model failure. Instead, we interpret this as a genuine \textit{signal} in the data: some nights show significantly more RFI contamination than others, beyond what would be expected from random sampling of a Bernoulli process, so the inferred season-level value of $w_\mathrm{RFI}$ depends sensitively on which nights happen to fall into each subset. Consistent with this interpretation, the discrepancies are symmetric across random splits, with either half yielding a larger $w_\mathrm{RFI}$ estimate depending on the particular realization.

\begin{figure}
    \centering
    \includegraphics[width=1.0\linewidth]{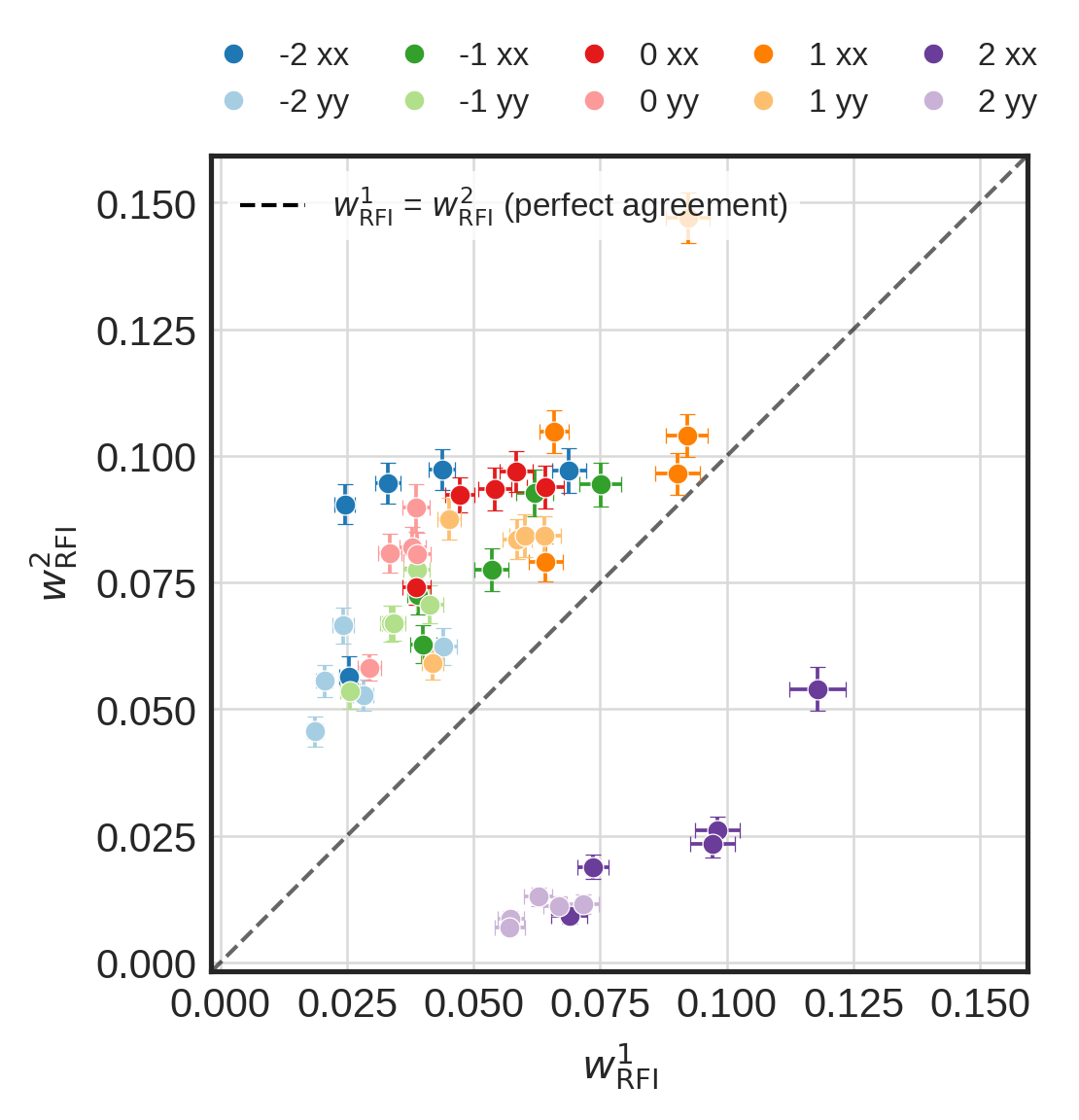}
    \caption{Comparison of the inferred seasonal RFI occupancy between the first and second chronological halves of the observing season, denoted $w_\mathrm{RFI}^1$ and $w_\mathrm{RFI}^2$, respectively. Colors denote different pointing and polarization combinations.}
    \label{fig:chrono_split}
\end{figure}

To further test this interpretation, we repeat the analysis using a \textit{chronological} rather than random division of the season.  The first subset contains the first chronological half of the observing nights, with inferred RFI occupancy $w_\mathrm{RFI}^1$, while the second contains the remainder, with inferred RFI occupancy $w_\mathrm{RFI}^2$. We again perform inference independently on the two subsets and compare the resulting parameters. In contrast to the random splits, the chronological division produces a clear systematic offset between the two halves. Figure~\ref{fig:chrono_split} illustrates this behavior. Here, the signal is obvious: the second half of the season contains more RFI contamination than the first half of the season for all pointings except pointing 2, where the opposite behavior is observed.

This temporal variation suggests that a single season-wide value of $w_\mathrm{RFI}$ may be too restrictive. A natural extension would be to allow these mixture weights to vary across the observing season, for example by assigning separate parameters to shorter temporal segments. This would reduce the pooling of information across periods with different contamination levels and allow the model to capture genuine changes in the RFI environment over the course of the season. We leave the implementation of such a time-dependent formulation to future work.

\begin{figure*}
    \centering
    \includegraphics[width=1.0\linewidth]{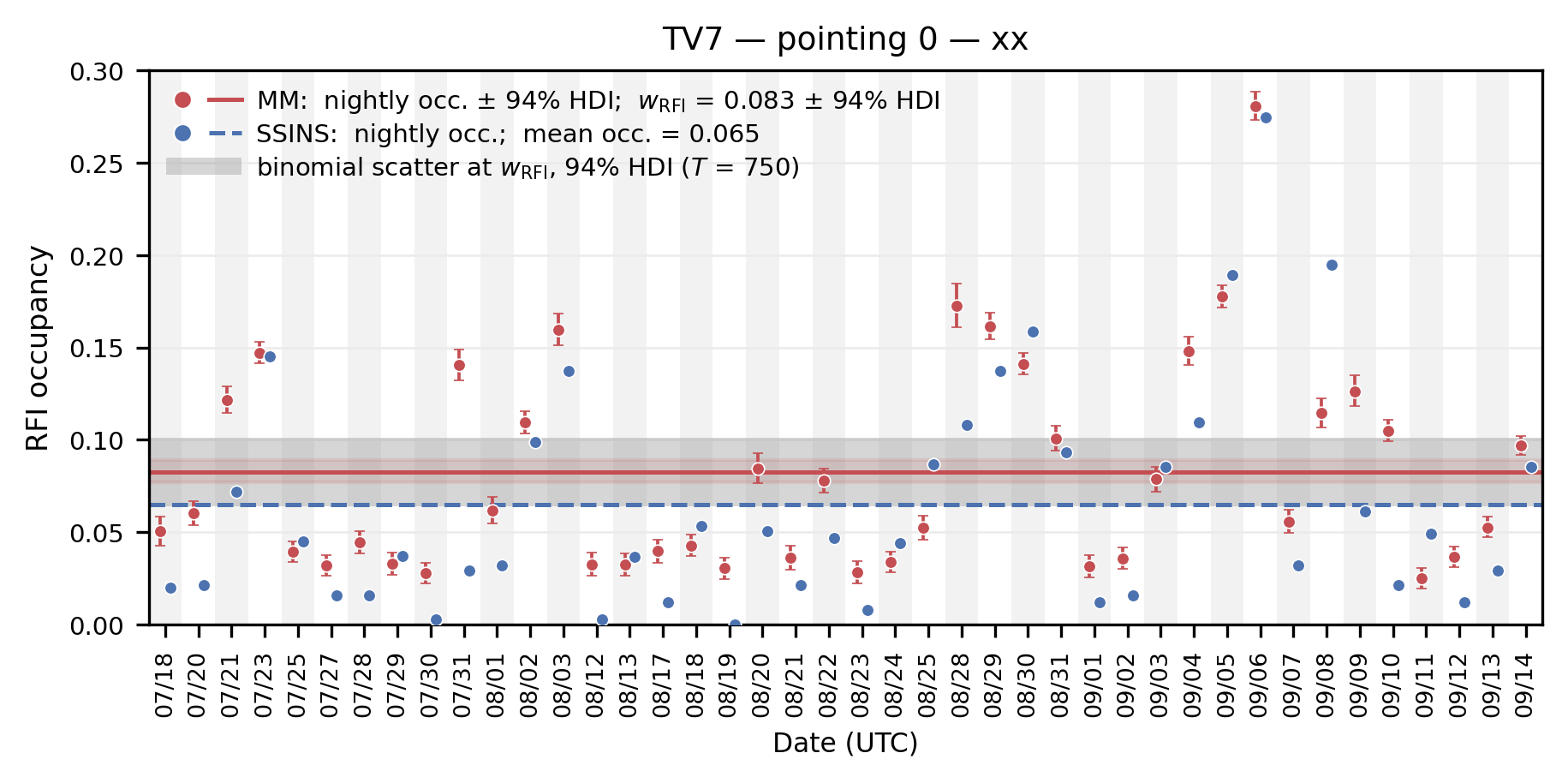}
    \caption{Comparison of nightly RFI occupancy inferred by the mixture model and estimated from \textsc{SSINS} flags for the TV7 frequency band, pointing 0, and XX polarization. In red: nightly mixture model occupancy, obtained by averaging the inferred soft labels over time. Error bars indicate the 94\% HDI obtained from the posterior distribution of the nightly averages. The red band shows the 94\% HDI of the season-level $w_\mathrm{RFI}$, while the gray band corresponds to binomial scatter (see Section~\ref{sec:model_checking}). In blue: nightly \textsc{SSINS} occupancy, obtained by averaging the binary SSINS flags over time after collapsing the frequency axis with a logical OR. The blue dashed line denotes the mean \textsc{SSINS} occupancy across the season. Dates are given in UTC in MM/DD format; all observations are from 2014. The vertical alternating shaded bands are used to help visually distinguish which points belong to which night.}
    \label{fig:compare_ssins}
\end{figure*}

To visualize the temporal variation in RFI contamination more directly, we perform a final model check based on the inferred soft labels. For each night, we average the soft labels over the time axis to obtain a nightly RFI occupancy. The resulting occupancies are shown in Figure~\ref{fig:compare_ssins}, together with the season-level inferred $w_\mathrm{RFI}$. The uncertainty on each nightly estimate is obtained by repeating this calculation for each posterior sample and then computing the 94\% HDI across chains and draws. For clarity, we show only the nights corresponding to the TV7, pointing 0, and XX polarization combination, while noting that the trends discussed below are qualitatively similar across the remaining combinations.

Figure~\ref{fig:compare_ssins} confirms the temporal variation suggested by the previous checks: nights in the second half of the season generally exhibit higher RFI occupancy. The variation is nevertheless more complex than a simple seasonal trend, as the nightly estimates appear to cluster around several distinct values of $w_\mathrm{RFI}$, suggesting multiple populations of RFI occupancy. A hierarchical extension of the model could account for this structure by allowing observations to switch between distinct occupancy states, for example through an additional mixture model over $w_\mathrm{RFI}$.

Most nightly occupancy estimates do not overlap the 94\% HDI of the season-level $w_\mathrm{RFI}$. This is not unexpected, because the two quantities describe different statistical properties of the data. The individual points measure the RFI occupancy of particular nights, whereas $w_\mathrm{RFI}$ characterizes the full season for the selected pointing, frequency band, and polarization. The small error bars on those points just mean that the model is confident that those nights are especially contaminated, or especially clean. The gray band provides a separate test of whether the observed night-to-night spread can be explained by finite sampling alone. If each time sample were independently contaminated with probability $w_\mathrm{RFI}$, then the occupancy of a night containing $T$ samples would follow $\mathrm{Binomial}(T,w_\mathrm{RFI})/T$. We estimate the 94\% HDI of this distribution by Monte Carlo using the median number of samples across nights, $T=750$. Most nights lie well outside this band, showing that the observed variation exceeds that expected from finite sampling alone and is therefore genuine, consistent with the signal already seen in Figures~\ref{fig:random_split} and~\ref{fig:chrono_split}.

We compare the inferred nightly occupancies with those obtained independently using \textsc{SSINS}, also shown in Figure~\ref{fig:compare_ssins}. Here, we specifically look at the \textsc{SSINS} flags, produced according to Section~\ref{sec:ssins} and collapsed to a one-dimensional time-series representation as described in Section~\ref{sec:specific_example}. Averaging these flags over time gives the nightly \textsc{SSINS} occupancy. The blue dashed line denotes the corresponding mean \textsc{SSINS} occupancy across all nights.

The mixture model and \textsc{SSINS} occupancies exhibit the same qualitative temporal behavior. In particular, nights assigned high RFI occupancy by the mixture model also tend to show elevated occupancy according to \textsc{SSINS}. This agreement provides an additional independent check that the mixture model is capturing genuine temporal variations in RFI contamination.

\subsection{Summary Statistics}\label{sec:summary_statistics}

\begin{figure*}
    \centering
    \includegraphics[width=1.0\linewidth]{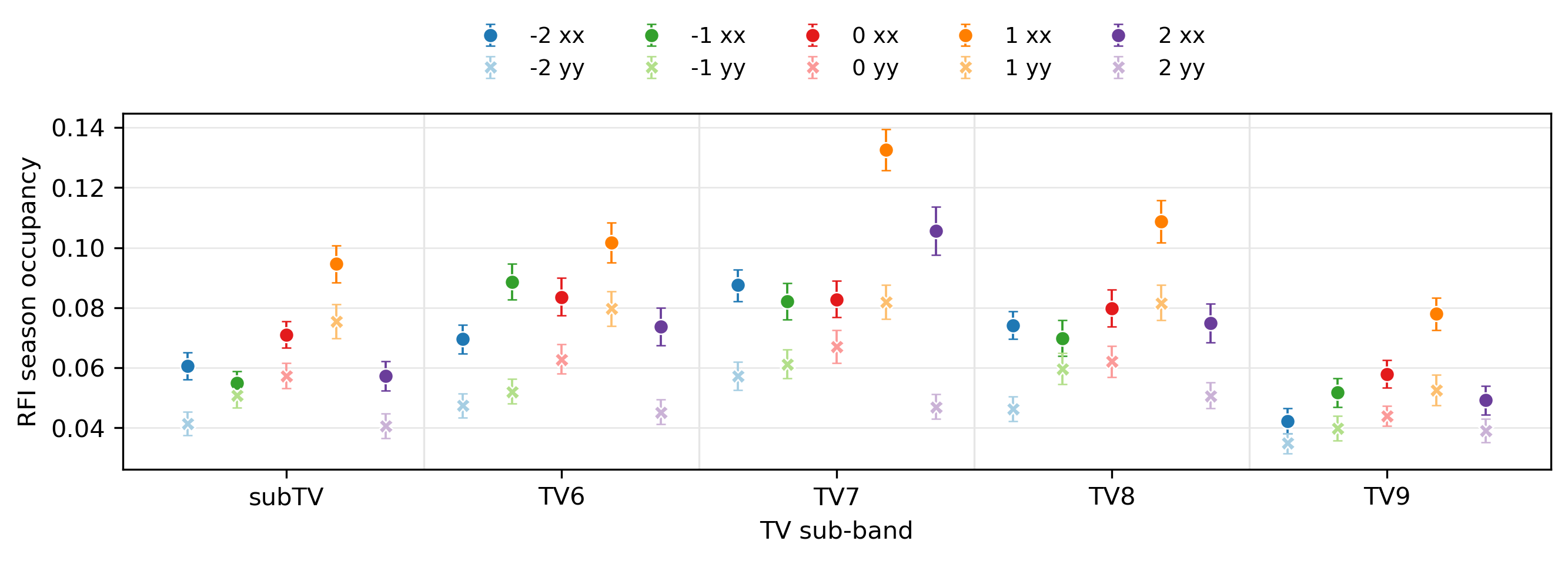}
    \caption{Inferred RFI occupancy as a function of frequency sub-band for all pointings and polarizations. Colors denote the different pointing and polarization combinations, and error bars indicate the 94\% HDIs of the posterior distributions. The x-axis shows the five frequency sub-bands considered in this study. Within each sub-band, horizontal offsets are introduced solely to visually separate the different pointings and have no physical meaning. XX and YY polarizations for a given pointing are plotted at the same horizontal position for clarity.}
    \label{fig:season_summary}
\end{figure*}

After running inference independently for all 50 combinations of pointing, frequency sub-band, and polarization, we can synthesize the resulting outputs to draw conclusions at the level of the full observing season. Figure~\ref{fig:season_summary} shows the inferred $w_\text{RFI}$ parameter, representing the RFI occupancy, for all pointing, frequency sub-band, and polarization combination, along with the corresponding 94\% HDI.

Several of these conclusions can also be obtained with other established flagging algorithms such as \textsc{SSINS}. Given that the 2014 season considered here was independently processed via 
\textsc{SSINS} as part of the data pre-processing, the flags obtained as a byproduct of that processing provide a useful basis for comparison and cross-validation of our mixture-model results. Among the trends recovered by both approaches, we highlight several qualitative features.

First, we consistently find the XX polarization to be more heavily contaminated than the YY polarization. Previous imaging studies at the Murchison Radio-astronomy Observatory have identified aircraft reflections as a source of RFI \citep{Tingay_2020, Ducharme_Pober_2025}, and \citet{Wilensky_2023} report that similar DTV events in SSINS are frequently associated with aircraft reflections. However, other propagation mechanisms, including tropospheric ducting, scattering, and meteor-trail reflections, may also contribute. Since our analysis does not localize individual RFI sources, we cannot conclusively attribute the observed polarization dependence to a particular mechanism.

Next, we find the TV7 frequency band (181--188 MHz) to be more consistently contaminated than the other frequency sub-bands considered here. This is consistent with previous studies of the MWA RFI environment, which identify Australian DTV channel 7 as the dominant source of DTV contamination. In a recent analysis of the MWA 2016 observing season, channel 7 was flagged in 5-6\% of the data, substantially more than the neighboring DTV channels, which were flagged in less than 1\% of the data \citep{Kunicki_Pober_2024}.


Thus far, the conclusions presented here could also have been obtained using \textsc{SSINS}; indeed, our parallel \textsc{SSINS} analysis yields the same qualitative trends. The major improvement of this work is instead the \textit{quantitative} measurement of RFI occupancy and characterization of RFI environment. Although a quantitative estimate of RFI occupancy can be constructed from \textsc{SSINS} by, for example, averaging flags across frequency bands, polarizations, or pointings, such estimates are ultimately derived from categorical flagging masks. Our method instead provides a Bayesian estimate of the RFI occupancy together with its associated error bars. Across all 50 pointing, frequency sub-band, and polarization combinations, we find that the average soft labels closely agree with the inferred $w_\mathrm{RFI}$, whereas the average binary labels obtained by thresholding the mixture-model responsibilities at 0.5 typically yield lower occupancies, illustrating the information lost when ambiguous classifications are reduced to binary decisions.

\section{Limitations and future improvements}\label{sec:limitations}

The framework developed in this work necessarily makes several simplifying assumptions that limit its sensitivity to certain forms of RFI. Most importantly, the analysis is performed after averaging the data over relatively broad frequency sub-bands\footnote{The observations considered in our case study span 768 fine frequency channels. Sub-band averaging therefore averages across approximately 150 fine frequency channels.}. This helps reduce the dimensionality of the problem and reduces the overall noise level, but it also reduces sensitivity to narrowband RFI that may occupy only a small fraction of a given sub-band. Given that narrowband RFI is among the most common types of RFI in MWA data, extending the analysis to finer frequency resolution, and ultimately the full frequency array, is a logical future improvement to this pipeline. Doing so would preserve more of the spectral structure of the contamination, although at the cost of substantially greater computational expense.

A second limitation arises from the use of a two-component mixture model consisting of a clean component and a single RFI component. In practice, RFI can originate from a variety of sources with different statistical and temporal characteristics, and these may not be well described by a single contaminated distribution. A richer mixture model containing multiple RFI components could therefore provide a more realistic representation of the data and potentially distinguish between qualitatively different classes of contamination, or between RFI and other anomalous behavior, as discussed in Section~\ref{sec:model_ambiguities}.

The current mixture formulation may also become less well constrained for nights that are almost entirely free of RFI. Because the model assumes that the data arise from both clean and contaminated components, the parameters describing the RFI distribution may be weakly constrained when very few samples are actually contaminated with RFI. For our specific use case, considering that RFI contamination is only becoming more and more ubiquitous, this specific limitation is not particularly problematic. For other applications to anomaly-detection problems, however, future work could investigate formulations that more naturally accommodate single-component observations, for example through more informative priors or hierarchical structure linking the anomalous parameters across nights.

Returning to our case study, further improvements could also be introduced at the post-processing stage. The probabilistic flags produced by the mixture model could be supplemented with heuristics designed specifically for downstream science applications, including power spectrum analyses. For example, flags could be extended along the frequency axis to reduce chromatic flagging structure that may otherwise bias a power spectrum analysis \citep{wilensky_2022}. Similarly, short unflagged gaps between closely separated contaminated time samples could be filled when they are likely to belong to the same RFI event.

Finally, the present framework treats the latent state of each time sample independently and therefore does not explicitly model the temporal persistence of RFI events. A natural extension would be to introduce temporal dependence; for example, through a hidden Markov model, in which the probability of a sample being assigned to a clean or contaminated state depends on the state at neighboring time steps. Similar approaches have been used in astronomical time-series analysis to identify transient events by explicitly modeling transitions between latent states and to infer both event duration and associated uncertainty \citep{Esquivel_2025}. Such a formulation could provide more temporally coherent RFI classifications, improve sensitivity to extended low-level contamination, and reduce isolated false-positive or false-negative assignments. The performance of such an extension could also be compared against established methods that exploit the temporal clustering of RFI, including SumThreshold \citep{Offringa_2010} and morphological filling algorithms \citep{Offringa_2012}.

\section{Conclusion}\label{sec:conclusion}

Radio-frequency interference remains a pervasive challenge for modern radio astronomy, particularly for high-precision experiments that are sensitive to even weak residual contamination. In this work, we use interferometric observations from the Murchison Widefield Array as a case study and reformulate RFI identification and characterization as a general anomaly detection problem. By compressing the original interferometric data into one-dimensional time series, we develop a flexible and generalizable framework for characterizing RFI.

Our model assumes that the observed time series consist of a smoothly varying background, observational noise, and anomalous contamination associated with RFI. The background is modeled using a sequence of Legendre polynomials, while the residuals relative to this background are described by a two-component mixture. Clean samples are modeled using a zero-centered normal distribution, while RFI-contaminated samples are modeled using a positively skewed Cauchy distribution. For each time sample, we then calculate the posterior probability of belonging to the RFI component, producing probabilistic soft classification labels.

We apply the framework to the 2014 MWA Phase I EoR highband observing season as a worked example, fitting the model independently for all 50 combinations of pointing, polarization, and frequency sub-band. We assess the reliability of the inference using standard convergence diagnostics, visual inspection of the posterior distributions and inferred backgrounds, and a series of internal checks. Random divisions of the observing season reveal small but systematic differences between independently inferred subsets, while a chronological division exposes genuine temporal evolution in the RFI environment. We further compare the resulting classifications against those produced by the established Sky-Subtracted Incoherent Noise Spectra pipeline (\textsc{SSINS}). The two approaches generally agree on bright contamination, while differences appear for weaker contamination on the boundary between clean and RFI.

At the season level, the inferred RFI occupancies reveal clear structure across the various fit combinations. Contamination is generally stronger in the XX polarization than in YY, with the TV7 frequency band exhibiting higher RFI occupancy than the remaining bands. We also find pointing 1 to be more highly contaminated than the other pointings. Substantial variation is additionally observed from night to night and across the observing season, suggesting that RFI contamination may not be adequately described by a single season-wide occupancy. We leave the implementation of a time-dependent occupancy parameter to future work.

The principal contribution of this work is not simply an alternative flagging algorithm, but a probabilistic framework in which RFI detection and characterization are performed jointly. The resulting soft classifications quantify the confidence associated with individual detections, while the inferred population parameters enable broader statistical studies of the contamination itself. Although demonstrated here using MWA EoR observations, the underlying formalism is general and can be adapted to other radio astronomy data sets, as well as to anomaly detection problems in one-dimensional time-series data more broadly. Future work may extend the model to accommodate additional contamination populations, sample the frequency axis more finely, and explicitly model the temporal dependence of RFI. As the radio sky becomes increasingly affected by human-made interference, continued advances in RFI identification, characterization, and statistical modeling will be essential for preserving our ability to recover faint astrophysical signals and safeguarding the next generation of precision radio experiments.

\begin{acknowledgments}

This work makes use of data obtained from Inyarrimanha Ilgari Bundara, the Murchison Radio-astronomy Observatory. We acknowledge the Wajarri Yamaji as the Traditional Owners and Native Title Holders of the land on which the observatory is located.

This research was enabled in part by support provided by the Center for Computation and Visualization at Brown University (\url{https://ccv.brown.edu/}),  Calcul Québec (\url{https://www.calculquebec.ca/}) and the Digital Research Alliance of Canada (\url{https://www.alliancecan.ca}).

A. Li acknowledges support from the McGill University Faculty of Science through the Science Undergraduate Research Awards (SURA) Program. MW acknowledges support from the CITA National Fellows Program and the TSI Fellows Program. A. Liu acknowledges support from the Natural Sciences and Engineering Research Council of Canada through the Discovery Grants Program and the Alliance International Program, as well as from the William Dawson Scholar Program at McGill University.

This research made use of open-source scientific software, including NumPy \citep{numpy}, SciPy \citep{scipy}, Matplotlib \citep{matplotlib}, JAX \citep{jax}, NumPyro \citep{numpyro}, and ArviZ \citep{arviz}.

\end{acknowledgments}

\bibliography{main}{}

@PHDTHESIS{Star2024,
       author = {{Star}, Pyxie},
        title = "{Digital Nonlinearities and the first Epoch of Reionization Power Spectrum Limit from MWA Phase III}",
       school = {University of Washington, Seattle},
         year = 2024,
        month = Oct,
}

@ARTICLE{Kolopanis2023,
       author = {{Kolopanis}, Matthew and {Pober}, Jonathan C. and {Jacobs}, Daniel C. and {McGraw}, Samantha},
        title = "{New EoR power spectrum limits from MWA Phase II using the delay spectrum method and novel systematic rejection}",
      journal = {\mnras},
         year = 2023,
        month = jun,
       volume = {521},
       number = {4},
        pages = {5120-5138},
          doi = {10.1093/mnras/stad845},
archivePrefix = {arXiv},
       eprint = {2210.10885},
 primaryClass = {astro-ph.CO},
       adsurl = {https://ui.adsabs.harvard.edu/abs/2023MNRAS.521.5120K}
}

@ARTICLE{Offringa2015,
       author = {{Offringa}, A.~R. and {Wayth}, R.~B. and {Hurley-Walker}, N. and {Kaplan}, D.~L. and {Barry}, N. and {Beardsley}, A.~P. and {Bell}, M.~E. and {Bernardi}, G. and {Bowman}, J.~D. and {Briggs}, F. and {Callingham}, J.~R. and {Cappallo}, R.~J. and {Carroll}, P. and {Deshpande}, A.~A. and {Dillon}, J.~S. and {Dwarakanath}, K.~S. and {Ewall-Wice}, A. and {Feng}, L. and {For}, B.-Q. and {Gaensler}, B.~M. and {Greenhill}, L.~J. and {Hancock}, P. and {Hazelton}, B.~J. and {Hewitt}, J.~N. and {Hindson}, L. and {Jacobs}, D.~C. and {Johnston-Hollitt}, M. and {Kapi{\'n}ska}, A.~D. and {Kim}, H.-S. and {Kittiwisit}, P. and {Lenc}, E. and {Line}, J. and {Loeb}, A. and {Lonsdale}, C.~J. and {McKinley}, B. and {McWhirter}, S.~R. and {Mitchell}, D.~A. and {Morales}, M.~F. and {Morgan}, E. and {Morgan}, J. and {Neben}, A.~R. and {Oberoi}, D. and {Ord}, S.~M. and {Paul}, S. and {Pindor}, B. and {Pober}, J.~C. and {Prabu}, T. and {Procopio}, P. and {Riding}, J. and {Udaya Shankar}, N. and {Sethi}, S. and {Srivani}, K.~S. and {Staveley-Smith}, L. and {Subrahmanyan}, R. and {Sullivan}, I.~S. and {Tegmark}, M. and {Thyagarajan}, N. and {Tingay}, S.~J. and {Trott}, C.~M. and {Webster}, R.~L. and {Williams}, A. and {Williams}, C.~L. and {Wu}, C. and {Wyithe}, J.~S. and {Zheng}, Q.},
        title = "{The Low-Frequency Environment of the Murchison Widefield Array: Radio-Frequency Interference Analysis and Mitigation}",
      journal = {\pasa},
         year = 2015,
        month = mar,
       volume = {32},
          eid = {e008},
        pages = {e008},
          doi = {10.1017/pasa.2015.7},
archivePrefix = {arXiv},
       eprint = {1501.03946},
 primaryClass = {astro-ph.IM},
       adsurl = {https://ui.adsabs.harvard.edu/abs/2015PASA...32....8O}
}

@ARTICLE{Offringa2023,
       author = {{Offringa}, A.~R. and {Adebahr}, B. and {Kutkin}, A. and {Adams}, E.~A.~K. and {Oosterloo}, T.~A. and {van der Hulst}, J.~M. and {D{\'e}nes}, H. and {Bassa}, C.~G. and {Lucero}, D.~L. and {Blok}, W.~J.~G. and {Hess}, K.~M. and {van Leeuwen}, J. and {Loose}, G.~M. and {Maan}, Y. and {Oostrum}, L.~C. and {Orr{\'u}}, E. and {Vohl}, D. and {Ziemke}, J.},
        title = "{An interference detection strategy for Apertif based on AOFlagger 3}",
      journal = {\aap},
         year = 2023,
        month = feb,
       volume = {670},
          eid = {A166},
        pages = {A166},
          doi = {10.1051/0004-6361/202245022},
archivePrefix = {arXiv},
       eprint = {2301.01562},
 primaryClass = {astro-ph.IM},
       adsurl = {https://ui.adsabs.harvard.edu/abs/2023A&A...670A.166O}
}

@ARTICLE{Barry2019,
       author = {{Barry}, N. and {Wilensky}, M. and {Trott}, C.~M. and {Pindor}, B. and {Beardsley}, A.~P. and {Hazelton}, B.~J. and {Sullivan}, I.~S. and {Morales}, M.~F. and {Pober}, J.~C. and {Line}, J. and {Greig}, B. and {Byrne}, R. and {Lanman}, A. and {Li}, W. and {Jordan}, C.~H. and {Joseph}, R.~C. and {McKinley}, B. and {Rahimi}, M. and {Yoshiura}, S. and {Bowman}, J.~D. and {Gaensler}, B.~M. and {Hewitt}, J.~N. and {Jacobs}, D.~C. and {Mitchell}, D.~A. and {Udaya Shankar}, N. and {Sethi}, S.~K. and {Subrahmanyan}, R. and {Tingay}, S.~J. and {Webster}, R.~L. and {Wyithe}, J.~S.~B.},
        title = "{Improving the Epoch of Reionization Power Spectrum Results from Murchison Widefield Array Season 1 Observations}",
      journal = {\apj},
         year = 2019,
        month = oct,
       volume = {884},
       number = {1},
          eid = {1},
        pages = {1},
          doi = {10.3847/1538-4357/ab40a8},
archivePrefix = {arXiv},
       eprint = {1909.00561},
 primaryClass = {astro-ph.IM},
       adsurl = {https://ui.adsabs.harvard.edu/abs/2019ApJ...884....1B}
}

@ARTICLE{mwa1,
       author = {{Tingay}, S.~J. and {Goeke}, R. and {Bowman}, J.~D. and {Emrich}, D. and {Ord}, S.~M. and {Mitchell}, D.~A. and {Morales}, M.~F. and {Booler}, T. and {Crosse}, B. and {Wayth}, R.~B. and {Lonsdale}, C.~J. and {Tremblay}, S. and {Pallot}, D. and {Colegate}, T. and {Wicenec}, A. and {Kudryavtseva}, N. and {Arcus}, W. and {Barnes}, D. and {Bernardi}, G. and {Briggs}, F. and {Burns}, S. and {Bunton}, J.~D. and {Cappallo}, R.~J. and {Corey}, B.~E. and {Deshpande}, A. and {Desouza}, L. and {Gaensler}, B.~M. and {Greenhill}, L.~J. and {Hall}, P.~J. and {Hazelton}, B.~J. and {Herne}, D. and {Hewitt}, J.~N. and {Johnston-Hollitt}, M. and {Kaplan}, D.~L. and {Kasper}, J.~C. and {Kincaid}, B.~B. and {Koenig}, R. and {Kratzenberg}, E. and {Lynch}, M.~J. and {Mckinley}, B. and {Mcwhirter}, S.~R. and {Morgan}, E. and {Oberoi}, D. and {Pathikulangara}, J. and {Prabu}, T. and {Remillard}, R.~A. and {Rogers}, A.~E.~E. and {Roshi}, A. and {Salah}, J.~E. and {Sault}, R.~J. and {Udaya-Shankar}, N. and {Schlagenhaufer}, F. and {Srivani}, K.~S. and {Stevens}, J. and {Subrahmanyan}, R. and {Waterson}, M. and {Webster}, R.~L. and {Whitney}, A.~R. and {Williams}, A. and {Williams}, C.~L. and {Wyithe}, J.~S.~B.},
        title = "{The Murchison Widefield Array: The Square Kilometre Array Precursor at Low Radio Frequencies}",
      journal = {\pasa},
         year = 2013,
        month = jan,
       volume = {30},
          eid = {e007},
        pages = {e007},
          doi = {10.1017/pasa.2012.007},
archivePrefix = {arXiv},
       eprint = {1206.6945},
 primaryClass = {astro-ph.IM},
       adsurl = {https://ui.adsabs.harvard.edu/abs/2013PASA...30....7T}
}

@ARTICLE{mwa2,
       author = {{Wayth}, Randall B. and {Tingay}, Steven J. and {Trott}, Cathryn M. and {Emrich}, David and {Johnston-Hollitt}, Melanie and {McKinley}, Ben and {Gaensler}, B.~M. and {Beardsley}, A.~P. and {Booler}, T. and {Crosse}, B. and {Franzen}, T.~M.~O. and {Horsley}, L. and {Kaplan}, D.~L. and {Kenney}, D. and {Morales}, M.~F. and {Pallot}, D. and {Sleap}, G. and {Steele}, K. and {Walker}, M. and {Williams}, A. and {Wu}, C. and {Cairns}, Iver. H. and {Filipovic}, M.~D. and {Johnston}, S. and {Murphy}, T. and {Quinn}, P. and {Staveley-Smith}, L. and {Webster}, R. and {Wyithe}, J.~S.~B.},
        title = "{The Phase II Murchison Widefield Array: Design overview}",
      journal = {\pasa},
         year = 2018,
        month = nov,
       volume = {35},
          eid = {e033},
        pages = {e033},
          doi = {10.1017/pasa.2018.37},
archivePrefix = {arXiv},
       eprint = {1809.06466},
 primaryClass = {astro-ph.IM},
       adsurl = {https://ui.adsabs.harvard.edu/abs/2018PASA...35...33W}
}

@article{AOFlagger, 
title={The Low-Frequency Environment of the Murchison Widefield Array: Radio-Frequency Interference Analysis and Mitigation},
volume={32}, 
DOI={10.1017/pasa.2015.7}, 
journal={Publications of the Astronomical Society of Australia}, 
author={Offringa, A. R. and Wayth, R. B. and Hurley-Walker, N. and Kaplan, D. L. and Barry, N. and Beardsley, A. P. and Bell, M. E. and Bernardi, G. and Bowman, J. D. and Briggs, F. and et al.}, 
year={2015}, 
pages={e008}}

@article{prabu2023,
author = {Prabu, Steve and Tingay, Steven and Williams, A.},
year = {2023},
month = {11},
pages = {1-10},
title = {A Near-Field Treatment of Aperture Synthesis Techniques using the Murchison Widefield Array},
volume = {40},
journal = {Publications of the Astronomical Society of Australia},
doi = {10.1017/pasa.2023.56}
}

@article{Wilensky_2019,
doi = {10.1088/1538-3873/ab3cad},
url = {https://dx.doi.org/10.1088/1538-3873/ab3cad},
year = {2019},
month = {oct},
publisher = {The Astronomical Society of the Pacific},
volume = {131},
number = {1005},
pages = {114507},
author = {Michael J. Wilensky and Miguel F. Morales and Bryna J. Hazelton and Nichole Barry and Ruby Byrne and Sumit Roy},
title = {Absolving the SSINS of Precision Interferometric Radio Data: A New Technique for Mitigating Faint Radio Frequency Interference},
journal = {Publications of the Astronomical Society of the Pacific}
}

@article{Wilensky_2023,
doi = {10.3847/1538-4357/acffbd},
url = {https://dx.doi.org/10.3847/1538-4357/acffbd},
year = {2023},
month = {nov},
publisher = {The American Astronomical Society},
volume = {957},
number = {2},
pages = {78},
author = {Michael J. Wilensky and Miguel F. Morales and Bryna J. Hazelton and Pyxie L. Star and Nichole Barry and Ruby Byrne and C. H. Jordan and Daniel C. Jacobs and Jonathan C. Pober and C. M. Trott},
title = {Evidence of Ultrafaint Radio Frequency Interference in Deep 21 cm Epoch of Reionization Power Spectra with the Murchison Wide-field Array},
journal = {The Astrophysical Journal}
}

@article{DiVruno_2023,
	author = {Di Vruno, F. and Winkel, B. and Bassa, C. G. and Józsa, G. I. G. and Brentjens, M. A. and Jessner, A. and Garrington, S.},
	title = {Unintended electromagnetic radiation from Starlink satellites detected with LOFAR between 110 and 188 MHz},
	DOI= "10.1051/0004-6361/202346374",
	url= "https://doi.org/10.1051/0004-6361/202346374",
	journal = {A\&A},
	year = 2023,
	volume = 676,
	pages = "A75",
}

@article{Wilensky_2020,
    author = {Wilensky, Michael J and Barry, Nichole and Morales, Miguel F and Hazelton, Bryna J and Byrne, Ruby},
    title = "{Quantifying excess power from radio frequency interference in Epoch of Reionization measurements}",
    journal = {Monthly Notices of the Royal Astronomical Society},
    volume = {498},
    number = {1},
    pages = {265-275},
    year = {2020},
    month = {08},
    issn = {0035-8711},
    doi = {10.1093/mnras/staa2442},
    url = {https://doi.org/10.1093/mnras/staa2442},
    eprint = {https://academic.oup.com/mnras/article-pdf/498/1/265/33705825/staa2442.pdf},
}

@article{Grigg_2023,
	author = {{Grigg, D.} and {Tingay, S. J.} and {Sokolowski, M.} and {Wayth, R. B.} and {Indermuehle, B.} and {Prabu, S.}},
	title = {Detection of intended and unintended emissions from Starlink satellites in the SKA-Low frequency range, at the SKA-Low site, with an SKA-Low station analogue},
	DOI= "10.1051/0004-6361/202347654",
	url= "https://doi.org/10.1051/0004-6361/202347654",
	journal = {A\&A},
	year = 2023,
	volume = 678,
	pages = "L6",
}

@article{Tingay_2020,
   title={A survey of spatially and temporally resolved radio frequency interference in the FM band at the Murchison Radio-astronomy Observatory},
   volume={37},
   ISSN={1448-6083},
   url={http://dx.doi.org/10.1017/pasa.2020.32},
   DOI={10.1017/pasa.2020.32},
   journal={Publications of the Astronomical Society of Australia},
   publisher={Cambridge University Press (CUP)},
   author={Tingay, S. J. and Sokolowski, M. and Wayth, R. and Ung, D.},
   year={2020} }

@article{Kunicki_Pober_2024, title={χ2 from redundant calibration as a tool in the detection of faint radio-frequency interference}, volume={41}, DOI={10.1017/pasa.2024.79}, journal={Publications of the Astronomical Society of Australia}, author={Kunicki, Theodora and Pober, Jonathan C.}, year={2024}, pages={e097}}

@article{Ducharme_Pober_2025, title={Altitude estimation of radio frequency interference sources via interferometric near-field corrections}, volume={42}, DOI={10.1017/pasa.2024.123}, journal={Publications of the Astronomical Society of Australia}, author={Ducharme, Jade M. and Pober, Jonathan C.}, year={2025}, pages={e010}}

@misc{dtv_allocations,
  author       = {{Australian Communications and Media Authority}},
  title        = {Digital Terrestrial Television Broadcasting Planning Handbook, including Technical and General Assumptions},
  year         = {2005},
  url          = {https://www.acma.gov.au/sites/default/files/2021-02/Digital_Terrestrial_Television_Broadcasting_Planning_Handbook_including_technical_and_general_assumptions.pdf},
  note         = {Accessed: 2025-10-23, p.~65}
}

@article{starlink3,
	author = {{Bassa, C. G.} and {Di Vruno, F.} and {Winkel, B.} and {Józsa, G. I. G.} and {Brentjens, M. A.} and {Zhang, X.}},
	title = {Bright unintended electromagnetic radiation from second-generation Starlink satellites},
	DOI= "10.1051/0004-6361/202451856",
	url= "https://doi.org/10.1051/0004-6361/202451856",
	journal = {A\&A},
	year = 2024,
	volume = 689,
	pages = "L10",
}

@article{wilensky_2022,
    author = {Wilensky, Michael J and Hazelton, Bryna J and Morales, Miguel F},
    title = {Exploring the consequences of chromatic data excision in 21-cm epoch of reionization power spectrum observations},
    journal = {Monthly Notices of the Royal Astronomical Society},
    volume = {510},
    number = {4},
    pages = {5023-5034},
    year = {2022},
    month = {03},
    issn = {0035-8711},
    doi = {10.1093/mnras/stab3456},
    url = {https://doi.org/10.1093/mnras/stab3456},
    eprint = {https://academic.oup.com/mnras/article-pdf/510/4/5023/42245595/stab3456.pdf},
}

@book{bishop2006,
  author    = {Bishop, Christopher M.},
  title     = {Pattern Recognition and Machine Learning},
  series    = {Information Science and Statistics},
  publisher = {Springer},
  address   = {New York},
  year      = {2006},
  isbn      = {978-0-387-31073-2},
}

@article{Grigg_2025, 
title={Enhanced detection and identification of satellites using an all-sky multi-frequency survey with prototype SKA-Low stations}, 
volume={42}, 
DOI={10.1017/pasa.2024.136}, 
journal={Publications of the Astronomical Society of Australia}, 
author={Grigg, Dylan and Tingay, Steven and Prabu, Steve and Sokolowski, Marcin and Indermuehle, Balthasar}, 
year={2025}, 
pages={e015} }

@phdthesis{barry_2018,
  author  = "Barry, Nichole",
  title   = "Enhancing EoR limits through improved instrumental calibration of the MWA",
  school  = "University of Washington",
  year    = "2018"
}

@article{Li_2019,
doi = {10.3847/1538-4357/ab55e4},
url = {https://doi.org/10.3847/1538-4357/ab55e4},
year = {2019},
month = {dec},
publisher = {The American Astronomical Society},
volume = {887},
number = {2},
pages = {141},
author = {Li, W. and Pober, J. C. and Barry, N. and Hazelton, B. J. and Morales, M. F. and Trott, C. M. and Lanman, A. and Wilensky, M. and Sullivan, I. and Beardsley, A. P. and Booler, T. and Bowman, J. D. and Byrne, R. and Crosse, B. and Emrich, D. and Franzen, T. M. O. and Hasegawa, K. and Horsley, L. and Johnston-Hollitt, M. and Jacobs, D. C. and Jordan, C. H. and Joseph, R. C. and Kaneuji, T. and Kaplan, D. L. and Kenney, D. and Kubota, K. and Line, J. and Lynch, C. and McKinley, B. and Mitchell, D. A. and Murray, S. and Pallot, D. and Pindor, B. and Rahimi, M. and Riding, J. and Sleap, G. and Steele, K. and Takahashi, K. and Tingay, S. J. and Walker, M. and Wayth, R. B. and Webster, R. L. and Williams, A. and Wu, C. and Wyithe, J. S. B. and Yoshiura, S. and Zheng, Q.},
title = {First Season MWA Phase II Epoch of Reionization Power Spectrum Results at Redshift 7},
journal = {The Astrophysical Journal}
}

@ARTICLE{gelman_rubin,
       author = {{Gelman}, Andrew and {Rubin}, Donald B.},
        title = "{Inference from Iterative Simulation Using Multiple Sequences}",
      journal = {Statistical Science},
         year = 1992,
        month = jan,
       volume = {7},
        pages = {457-472},
          doi = {10.1214/ss/1177011136},
       adsurl = {https://ui.adsabs.harvard.edu/abs/1992StaSc...7..457G}
}

@misc{indermuehle2026,
      title={Tropospheric Ducting Prediction based on GFS Model Data at Inyarrimanha Ilgari Bundara, the CSIRO Murchison Radio-astronomy Observatory}, 
      author={Balthasar Indermuehle and Hajime Suzuki},
      year={2026},
      eprint={2608.12861},
      archivePrefix={arXiv},
      primaryClass={astro-ph.IM},
      url={https://arxiv.org/abs/2608.12861}, 
}

@misc{mwa_sweetspots,
  author       = {Williams, Andrew},
  title        = {MWA `sweet spots' and gridpoint numbers},
  howpublished = {MWA Telescope Documentation},
  year         = {2024},
  note         = {Updated May 16, 2024},
  url          = {https://mwatelescope.atlassian.net/wiki/spaces/MP/pages/24969505/MWA+sweet+spots+and+gridpoint+numbers}
}

@article{Esquivel_2025,
   title={Detecting Stellar Flares in Photometric Data Using Hidden Markov Models},
   volume={979},
   ISSN={1538-4357},
   url={http://dx.doi.org/10.3847/1538-4357/ad95f6},
   DOI={10.3847/1538-4357/ad95f6},
   number={2},
   journal={The Astrophysical Journal},
   publisher={American Astronomical Society},
   author={Esquivel, J. Arturo and Shen, Yunyi and Leos-Barajas, Vianey and Eadie, Gwendolyn and Speagle, Joshua S. and Craiu, Radu V and Medina, Amber and Davenport, James R. A.},
   year={2025},
   month=Jan, pages={141} }

@article{Nunhokee_2025,
doi = {10.3847/1538-4357/adda45},
url = {https://doi.org/10.3847/1538-4357/adda45},
year = {2025},
month = {aug},
publisher = {The American Astronomical Society},
volume = {989},
number = {1},
pages = {57},
author = {Nunhokee, C. D. and Null, D. and Trott, C. M and Barry, N. and Qin, Y. and Wayth, R. B. and Line, J. L. B. and Jordan, C. H. and Pindor, B. and Cook, J. H. and Bowman, J. and Chokshi, A. and Ducharme, J. and Elder, K. and Guo, Q. and Hazelton, B. and Hidayat, W. and Ito, T. and Jacobs, D. and Jong, E. and Kolopanis, M. and Kunicki, T. and Lilleskov, E. and Morales, M. F. and Pober, J. C. and Selvaraj, A. and Shi, R. and Takahashi, K. and Tingay, S. J. and Webster, R. L. and Yoshiura, S. and Zheng, Q.},
title = {Limits on the 21 cm Power Spectrum at z = 6.5–7.0 from Murchison Widefield Array Observations},
journal = {The Astrophysical Journal}
}

@article{Offringa_2010,
   title={Post-correlation radio frequency interference classification methods},
   ISSN={1365-2966},
   url={http://dx.doi.org/10.1111/j.1365-2966.2010.16471.x},
   DOI={10.1111/j.1365-2966.2010.16471.x},
   journal={Monthly Notices of the Royal Astronomical Society},
   publisher={Oxford University Press (OUP)},
   author={Offringa, A. R. and de Bruyn, A. G. and Biehl, M. and Zaroubi, S. and Bernardi, G. and Pandey, V. N.},
   year={2010},
   month=Mar }

@article{Offringa_2012,
   title={A morphological algorithm for improving radio-frequency interference detection},
   volume={539},
   ISSN={1432-0746},
   url={http://dx.doi.org/10.1051/0004-6361/201118497},
   DOI={10.1051/0004-6361/201118497},
   journal={Astronomy \& Astrophysics},
   publisher={EDP Sciences},
   author={Offringa, A. R. and van de Gronde, J. J. and Roerdink, J. B. T. M.},
   year={2012},
   month=Feb, pages={A95} }

@article{Offringa_2013,
   title={The brightness and spatial distributions of terrestrial radio sources},
   volume={435},
   ISSN={0035-8711},
   url={http://dx.doi.org/10.1093/mnras/stt1337},
   DOI={10.1093/mnras/stt1337},
   number={1},
   journal={Monthly Notices of the Royal Astronomical Society},
   publisher={Oxford University Press (OUP)},
   author={Offringa, A. R. and de Bruyn, A. G. and Zaroubi, S. and Koopmans, L. V. E. and Wijnholds, S. J. and Abdalla, F. B. and Brouw, W. N. and Ciardi, B. and Iliev, I. T. and Harker, G. J. A. and Mellema, G. and Bernardi, G. and Zarka, P. and Ghosh, A. and Alexov, A. and Anderson, J. and Asgekar, A. and Avruch, I. M. and Beck, R. and Bell, M. E. and Bell, M. R. and Bentum, M. J. and Best, P. and Bîrzan, L. and Breitling, F. and Broderick, J. and Brüggen, M. and Butcher, H. R. and de Gasperin, F. and de Geus, E. and de Vos, M. and Duscha, S. and Eislöffel, J. and Fallows, R. A. and Ferrari, C. and Frieswijk, W. and Garrett, M. A. and Grießmeier, J. and Hassall, T. E. and Horneffer, A. and Iacobelli, M. and Juette, E. and Karastergiou, A. and Klijn, W. and Kondratiev, V. I. and Kuniyoshi, M. and Kuper, G. and van Leeuwen, J. and Loose, M. and Maat, P. and Macario, G. and Mann, G. and McKean, J. P. and Meulman, H. and Norden, M. J. and Orru, E. and Paas, H. and Pandey-Pommier, M. and Pizzo, R. and Polatidis, A. G. and Rafferty, D. and Reich, W. and van Nieuwpoort, R. and Röttgering, H. and Scaife, A. M. M. and Sluman, J. and Smirnov, O. and Sobey, C. and Tagger, M. and Tang, Y. and Tasse, C. and Veen, S. ter and Toribio, C. and Vermeulen, R. and Vocks, C. and van Weeren, R. J. and Wise, M. W. and Wucknitz, O.},
   year={2013},
   month=Aug, pages={584–596} }

@article{hoffman2014,
  title   = {The No-U-Turn Sampler: Adaptively Setting Path Lengths in Hamiltonian Monte Carlo},
  author  = {Hoffman, Matthew D. and Gelman, Andrew},
  journal = {Journal of Machine Learning Research},
  volume  = {15},
  number  = {47},
  pages   = {1593--1623},
  year    = {2014}
}

@ARTICLE{scipy,
  author  = {Virtanen, Pauli and Gommers, Ralf and Oliphant, Travis E. and
            Haberland, Matt and Reddy, Tyler and Cournapeau, David and
            Burovski, Evgeni and Peterson, Pearu and Weckesser, Warren and
            Bright, Jonathan and {van der Walt}, St{\'e}fan J. and
            Brett, Matthew and Wilson, Joshua and Millman, K. Jarrod and
            Mayorov, Nikolay and Nelson, Andrew R. J. and Jones, Eric and
            Kern, Robert and Larson, Eric and Carey, C J and
            Polat, {\.I}lhan and Feng, Yu and Moore, Eric W. and
            {VanderPlas}, Jake and Laxalde, Denis and Perktold, Josef and
            Cimrman, Robert and Henriksen, Ian and Quintero, E. A. and
            Harris, Charles R. and Archibald, Anne M. and
            Ribeiro, Ant{\^o}nio H. and Pedregosa, Fabian and
            {van Mulbregt}, Paul and {SciPy 1.0 Contributors}},
  title   = {{{SciPy} 1.0: Fundamental Algorithms for Scientific
            Computing in Python}},
  journal = {Nature Methods},
  year    = {2020},
  volume  = {17},
  pages   = {261--272},
  adsurl  = {https://rdcu.be/b08Wh},
  doi     = {10.1038/s41592-019-0686-2},
}

@ARTICLE{numpy,
  author  = {Harris, Charles R. and Millman, K. Jarrod and
            van der Walt, Stéfan J and Gommers, Ralf and
            Virtanen, Pauli and Cournapeau, David and
            Wieser, Eric and Taylor, Julian and Berg, Sebastian and
            Smith, Nathaniel J. and Kern, Robert and Picus, Matti and
            Hoyer, Stephan and van Kerkwijk, Marten H. and
            Brett, Matthew and Haldane, Allan and
            Fernández del Río, Jaime and Wiebe, Mark and
            Peterson, Pearu and Gérard-Marchant, Pierre and
            Sheppard, Kevin and Reddy, Tyler and Weckesser, Warren and
            Abbasi, Hameer and Gohlke, Christoph and
            Oliphant, Travis E.},
  title   = {Array programming with {NumPy}},
  journal = {Nature},
  year    = {2020},
  volume  = {585},
  pages   = {357–362},
  doi     = {10.1038/s41586-020-2649-2}
}

@article{arviz,
doi = {10.21105/joss.09889},
url = {https://doi.org/10.21105/joss.09889},
year = {2026},
publisher = {The Open Journal},
volume = {11},
number = {119},
pages = {9889},
author = {Martin, Osvaldo A. and Abril-Pla, Oriol and Deklerk, Jordan and Axen, Seth D. and Carroll, Colin and Hartikainen, Ari and Vehtari, Aki},
title = {ArviZ: a modular and flexible library for exploratory analysis of Bayesian models},
journal = {Journal of Open Source Software}}

@article{matplotlib,
    author  = {Hunter, John D.},
    title   = {Matplotlib: A 2D Graphics Environment},
    journal = {Computing in Science \& Engineering},
    volume  = {9},
    number  = {3},
    pages   = {90--95},
    year    = {2007},
    doi     = {10.1109/MCSE.2007.55}
}

@software{jax,
  author = {James Bradbury and Roy Frostig and Peter Hawkins and Matthew James Johnson and Yash Katariya and Chris Leary and Dougal Maclaurin and George Necula and Adam Paszke and Jake Vander{P}las and Skye Wanderman-{M}ilne and Qiao Zhang},
  title = {{JAX}: composable transformations of {P}ython+{N}um{P}y programs},
  url = {http://github.com/jax-ml/jax},
  version = {0.3.13},
  year = {2018},
}

@misc{numpyro,
      title={Composable Effects for Flexible and Accelerated Probabilistic Programming in NumPyro}, 
      author={Du Phan and Neeraj Pradhan and Martin Jankowiak},
      year={2019},
      eprint={1912.11554},
      archivePrefix={arXiv},
      primaryClass={stat.ML},
      url={https://arxiv.org/abs/1912.11554}, 
}

@article{Leeney_2023,
   title={Bayesian approach to radio frequency interference mitigation},
   volume={108},
   ISSN={2470-0029},
   url={http://dx.doi.org/10.1103/PhysRevD.108.062006},
   DOI={10.1103/physrevd.108.062006},
   number={6},
   journal={Physical Review D},
   publisher={American Physical Society (APS)},
   author={Leeney, S. A. K. and Handley, W. J. and Acedo, E. de Lera},
   year={2023},
   month=Sept }

@misc{Leeney_2026,
      title={Scalable Bayesian data curation for next-generation radio experiments}, 
      author={S. A. K. Leeney and E. de Lera Acedo and W. J. Handley and H. T. J. Bevins and G. Allen and D. Anstey and K. Artuc and G. Bernardi and M. Bucher and S. Carey and J. Cavillot and R. Chiello and A. S. Chu and W. Croukamp and J. Cumner and S. Dasgupta and A. K. Dash and D. I. L. de Villiers and J. Dhandha and A. Dragovic and J. A. Ely and A. Fialkov and T. Gessey-Jones and C. Kirkham and G. Kulkarni and A. Magro and P. D. Meerburg and S. Mittal and D. Molnar and R. S. Patel and J. H. N. Pattison and S. Pegwal and C. M. Pieterse and J. R. Pritchard and G. M. Z. Rajpoot and N. Razavi-Ghods and D. Robins and I. L. V. Roque and A. Saxena and K. H. Scheutwinkel and E. Shen and P. H. Sims and M. Spinelli and J. L. Tutt and J. Zhu},
      year={2026},
      eprint={2606.16525},
      archivePrefix={arXiv},
      primaryClass={astro-ph.IM},
      url={https://arxiv.org/abs/2606.16525}, 
}
\bibliographystyle{aasjournalv7.1}



\end{document}